\documentclass[aps,pra,preprint,groupedaddress]{revtex4-2}

\usepackage{amssymb}
\usepackage{amsmath}
\usepackage{epsfig,graphicx,physics}

\usepackage{subfigure}
\usepackage{xcolor}

\begin{document}


\title{Quartic soliton solutions of a normal dispersion based mode-locked laser}


\author{M. Facão}
\email[]{mfacao@ua.pt}
\affiliation{Departament d’Enginyeria Electrònica, Universitat Politècnica de Catalunya, Terrassa 08222, Spain}
\affiliation{i3N and Departamento de Física, Universidade de Aveiro, Campus Universitário de Santiago, 3810-193 Aveiro, Portugal}

\author{D. Malheiro}
\email[]{diogomalheiro@ua.pt}
\affiliation{Instituto de Telecomunicações and i3N, Universidade de Aveiro, Campus Universitário de Santiago, 3810-193 Aveiro, Portugal}

\author{M. I. Carvalho}
\email[]{mines@fe.up.pt}
\affiliation{INESC TEC, Rua Dr.~Roberto Frias, 4200-465 Porto, Portugal}
\affiliation{Faculdade de Engenharia, Universidade do Porto, Rua Dr.~Roberto Frias, 4200-465 Porto, Portugal}

\date{\today}

\begin{abstract}
We studied the characteristics, regions of existence and stability of different types of solitons for a distributed model of a mode-locked laser whose dispersion is purely quartic and normal. Among the different types of solitons, we identified three main branches that are named according to their different amplitude: low, medium and high amplitude solitons. It was found that the first solitons are always unstable while the latter two exist and are stable in relatively large regions of the parameter space. Moreover, the stability regions of medium and high amplitude solitons overlap over a certain range of parameters, manifesting effects of bistability. The energy of high amplitude solitons increases quadratically with their width, whereas the energy of medium amplitude solitons may decrease or increase with the width depending on the parameter region. Furthermore, we have investigated the long term evolution of the continuous wave solutions under modulational instability, showing that medium amplitude solitons can arise in this scenario. Additionally, we assessed the effects of second and third order dispersion on medium and high amplitude solitons and found that both remain stable in the presence of these terms.\end{abstract}


\maketitle

\section{Introduction}
Quartic solitons are solitons of optical models whose dispersion is dominated by a fourth order term, whereas the term pure quartic solitons is reserved for models where only the fourth order dispersion is considered. They have been predicted \cite{karlsson94,akhmediev94,buryak95} and observed \cite{roy2013,blanco-redondo16} in conservative models but they have been more acclaimed in dissipative models for their advantages in mode-locked lasers and Kerr soliton frequency combs whenever the fourth order dispersion (4OD) is negative (anomalous 4OD). In fact, in mode-locked lasers, the negative quartic dispersion may give rise to pulses whose energy scales inversely with the width cubed, enabling highly energetic ultrashort pulses \cite{runge20,malheiro23,zhang24}. For Kerr combs presenting negative 4OD, the same energy-width relation was found but more importantly, the soliton spectra are flatter \cite{Taheri2019}. Quartic solitons have also been studied in more complex models like vectorial models for mode-locked lasers  \cite{he24} and a fiber resonator 
model allowing orbital angular momentum carrying modes \cite{shi24}.
In case of positive 4OD (also named normal 4OD), solitons do not exist for conservative models based on Kerr nonlinearity \cite{tam20} but do exist in dissipative models. A triangular-shaped pulse with a double peak spectrum has been reported by Runge \textit{et al.} \cite{runge20OL} for a modified nonlinear Schr\"odinger equation (NLSE) comprising positive 4OD, Kerr effect and gain as well as for a mode-locked laser lumped model. Note that, this kind of pulses was already reported as non-soliton solutions of the NLSE with positive second order dispersion (2OD) and positive 4OD \cite{bale11}. Another type of soliton solution for mode-locked laser models with positive 4OD was reported in \cite{quian22,malheiro23} which also presents a double-peaked spectrum but whose temporal profile consists of a sech central part on top of a large pedestal. Recently, and while we were obtaining the results here reported, the complex Ginzburg-Landau equation (CGLE) was used to obtain e\-xis\-ten\-ce regions for the solutions referred above \cite{wang24} and simulation and experimental results reported the existence of pulses in a fiber laser dominated by positive 4OD \cite{wu24}. In \cite{quian22}, a third kind of solution, that is asymmetrical, was reported. Nevertheless, all the reported solutions of mode-locked lasers dominated by normal 4OD were not yet sufficiently studied in terms of pulse characteristics, energy-width relation, regions of existence and stability, coexistence and onset from continuous wave evolution.

Here, we report a thorough study of soliton solutions of a distributed model for mode-locked lasers \cite{zaviyalov10,malheiro23} dominated by normal fourth order dispersion. Two of the solutions are similar to the above referred solutions, namely, the triangular-shaped and the sech-pulse above a pedestal. We have found that they coexist in some region of the parameter space, at which they are also stable, manifesting dynamics of bistability. The other ones were obtained by solving the associated ordinary differential equation but are shown to be unstable. Three main branches of those solutions (one always unstable and two stable in some regions) are characterized in terms of amplitude, phase and spectrum profiles and in terms of energy exchange with the exterior and within themselves. The other branches occur on a bifurcation region of parameters and only the amplitude profiles are presented.

The paper is organized as follows. The mathematical and theoretical framework for the distributed model is presented in Section \ref{sec:model_methods}, alongside a description of the numerical methods used to study the model. Section \ref{sec:solitons} characterizes the obtained soliton solutions. Their internal energy flow was studied, parameter regions of existence and stability were found and the energy-width scaling of stable solutions was analyzed. Section \ref{sec:modulation_instability} studies the formation of these solitons through modulational instability. In Section \ref{2&3_influence}, the effects of second order dispersion (2OD) and third order dispersion (3OD) on the existence and stability of the stable solitons is investigated. Section \ref{sec:conclusion} presents the main conclusions of the work.

\section{Model equations and methods} \label{sec:model_methods}
The evolution of the pulses in mode-locked lasers with a saturable absorber may be studied by a distributed model given by the following partial differential equation (PDE) \cite{zaviyalov10, malheiro23}
\begin{multline}
i\frac{\partial W}{\partial z}-\frac{1}{2}\left(\beta_2+ig_0T_2^2\right)\frac{\partial^2W}{\partial t^2}+i\frac{\beta_3}{6}\frac{\partial^3W}{\partial t^3}+\frac{\beta_4}{24}\frac{\partial^4W}{\partial t^4}= \\
i\left(\frac{g_0-k_\text{OC}/L}{2}\right)W-\frac{i}{2}\frac{d_\text{SA}/L}{1+|W|^2/\bar{P}_\text{sat}}W-\bar{\gamma}|W|^2W, 
\label{PDE}
\end{multline}
where $t$ is the retarded time, $z$ is the propagation distance, $W(z,t)$ is the slowly varying pulse envelope, $\beta_2$, $\beta_3$ and $\beta_4$ are the second, third  and fourth order dispersion parameters, $g_0$ is the small signal gain, $T_2$ is the inverse linewidth of the parabolic gain, $k_\text{OC}$ represents the losses of the output coupler, $L$ is the cavity length and $d_\text{SA}$ is the modulation depth of the saturable absorber. The parameters $\bar{\gamma}$ and $\bar{P}_\text{sat}$ are parameters associated with the nonlinear parameter $\gamma$ and saturation power $P_\text{sat}$ of the saturable absorber, respectively, and given by $\bar{\gamma}=\gamma(\exp(g_0L)-1)/g_0L$, $\bar{P}_\text{sat}=P_\text{sat}\exp(-g_0L)$. Apart from section \ref{2&3_influence}, the pulse solutions presented here are for $\beta_2=\beta_3=0$ and $\beta_4>0$, i.e., normal pure quartic solitons. To reduce the number of parameters, we derived a dimensionless equation, valid for this case, by performing the following change of variables
\begin{equation}
	q = \left(\frac{\bar{\gamma}}{a}\right)^{1/2}W, \quad Z = az, \quad T = \left(\frac{2a}{g_0T_2^2}\right)^{1/2}t,
\end{equation}
where $a = -g_0/2 + k_\text{OC}/2L + d_\text{SA}/2L$. Thus, the equation reads, 
\begin{equation}
	i\frac{\partial q}{\partial Z} + \frac{D_4}{24}\frac{\partial^4 q}{\partial T^4} + |q|^2q = i\alpha q + i\frac{\partial^2 q}{\partial T^2} - i\frac{\alpha + 1}{1+\rho|q|^2}q,
	\label{PDE2}
\end{equation}
having three dimensionless parameters given by
\begin{equation}
	D_4 = \frac{4\beta_4 a}{g_0^2T_2^4}, \quad \alpha = \frac{1}{a}\left(\frac{g_0}{2} - \frac{k_\text{OC}}{2L}\right), \quad \rho = \frac{a}{\bar{P}_\text{sat}\bar{\gamma}}.
\end{equation}
Applying the similarity variable transformation given by
$q(Z,T) = F(T)e^{i\sigma Z}$, where $F$ is a complex amplitude and $\sigma$ a real propagation constant, we obtain the following ordinary differential equation (ODE)
\begin{equation}
     \frac{D_4}{24}\frac{\partial^4 F}{\partial T^4}-i\frac{\partial^2 F}{\partial T^2}-\sigma F+|F|^2F=i\alpha F-i(1+\alpha)\frac{F}{1+\rho|F|^2}.     
     \label{ODE}
\end{equation}

Our search for pulse solutions has used two different methods: full integration of the Eq.~(\ref{PDE2}) using a localized input and integration of the ODE (\ref{ODE}).
The PDE in Eq.~(\ref{PDE2}) was solved through pseudo-spectral methods and the ODE in Eq.~(\ref{ODE}) was integrated using the Newton conjugate-gradient (NCG) method developed by J. Yang \cite{Yang2009, Yang2015}, starting from hyperbolic secant profiles or from soliton solutions obtained in previous simulations with similar parameters.

We have assumed that the pulses found by full integration of the PDE are stable since they survive to numerical error. However, the pulses found by the integration of the ODE could be unstable. To assess the stability of the latter ones we relied on two methods: integration of the PDE using a perturbed version of the ODE pulse as an input and observation of its evolution and calculation of the eigenvalues of the stability equations. For the latter method, the solution of Eq.~(\ref{PDE2}) was written as the solution of Eq.~(\ref{ODE}) plus a small perturbation  $\eta(Z, T)$, being 
$q(Z, T) = \left[F(T) + \eta(Z,T)\right]e^{i\sigma z}$ which, in first order, gives 
\begin{equation}
    i\eta_Z + \textbf{K}_{11}(F)\eta + \textbf{K}_{12}(F)\eta^\ast = 0
    \label{eq:perturb_evolution}
\end{equation}
where the $\ast$ denotes the complex conjugate and $\textbf{K}_{11}(F)$ and $\textbf{K}_{12}(F)$ are operators given by
\begin{align}
	\textbf{K}_{11}(F) &= \frac{D_4}{24}\partial_{T}^4 - i\partial_{T}^2 - \sigma - i\alpha + 2|F|^2 +  i\frac{1+\alpha}{\left(1+\rho|F|^2\right)^2}\\
	\textbf{K}_{12}(F) &= F^2 - i\frac{\left(1+\alpha\right)\rho F^2}{\left(1+\rho|F|^2\right)^2}.
	\label{eq:Kterms}
\end{align}
Considering that the evolution of the perturbation in $Z$ is exponential, namely, $\eta(Z, T) = v(T)e^{i\lambda Z} + w^\ast(T)e^{-i\lambda^\ast Z}$, we obtain the following eigenvalue problem
\begin{equation}
	\begin{bmatrix}
		\textbf{K}_{11} & \textbf{K}_{12}\\
		-\textbf{K}_{12}^\ast & -\textbf{K}_{11}^\ast
	\end{bmatrix}
	\begin{bmatrix}
		v \\ w
	\end{bmatrix}
	= \lambda
	\begin{bmatrix}
		v \\ w
	\end{bmatrix}.
	\label{stab}
\end{equation}
Note that if the imaginary part of $\lambda$, $\lambda_i$, is such that $\lambda_i > 0$, the perturbation will decay along $Z$ but if $\lambda_i < 0$, it will grow exponentially and the solution will become unstable. The eigenvalues were obtained by calculating the eigenvalues of the algebraic equations that result from the discretization of Eq.~\eqref{stab} using finite differences.

\section{Soliton solutions}\label{sec:solitons}
This section analyses the quartic soliton solutions of Eq. \eqref{PDE}. The first subsection presents pulse profiles and chirp of three types of solitons that were obtained. Subsection \ref{sec:energy_flow} studies the energy exchange dynamics of those solitons. In subsection \ref{sec:stability}, regions of existence and stability of those solitons are thoroughly discussed, and behaviors of bistability and hysteris are identified. Other types of soliton solutions that were not thoroughly characterized are also presented. Finally, subsection \ref{sec:ew} studies the energy-width scaling of stable solitons through the variation of equation parameters.

\subsection{Types of Solutions} \label{sec:types_of_sol}
The application of the methods referred in the previous section allowed us to find different types of soliton solutions of the model in Eq.~(\ref{PDE}) with null 2OD and 3OD and normal 4OD. Three main solution branches are studied more carefully and are named according to their relative peak amplitude as low amplitude soliton (LAS), medium amplitude soliton (MAS) and high amplitude soliton (HAS).
We have kept fixed $d_\text{SA} = 0.3$, $\gamma = 0.005$~W$^{-1}$m$^{-1}$, $L = 1$ m and $k_\text{OC} = -\text{ln}(0.3)$ as well as the saturation power $P_\text{sat}=80$~W and the 4OD coefficient $\beta_4=0.08$~ps$^4$m$^{-1}$.
This choice of parameters was based on previous works such as \cite{zaviyalov10} where the distributed model (with 2OD only) was introduced and \cite{quian22} which reported normal quartic solitons in mode-locked lasers. In an experimental setting, many of these parameters are intrinsic to the characteristics of the laser itself, be it the setup or its materials. However, effective dispersion management is fundamental for the generation of quartic solitons. The first quartic soliton fiber laser \cite{runge20} used an intracavity spectral pulse shaper based on a spatial light modulator to make $\beta_2$ and $\beta_3$ negligible while setting a value of negative $\beta_4$. Recently, the same technique was used to generate dissipative quartic solitons in a laser cavity with $\beta_4 > 0$ \cite{wu24}.

\begin{figure}[hbt!]
    \subfigure[]{
        \centering\includegraphics[height=0.25\textheight]{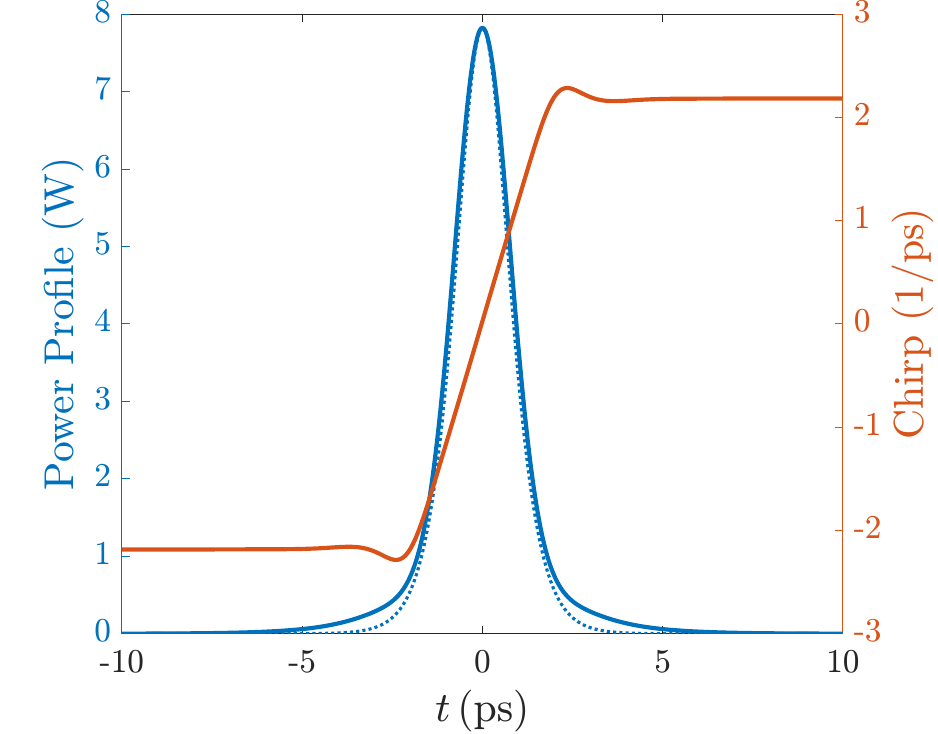}
        \label{LAS-PowerChirp}}
    \subfigure[]{
        \centering\includegraphics[height=0.25\textheight]{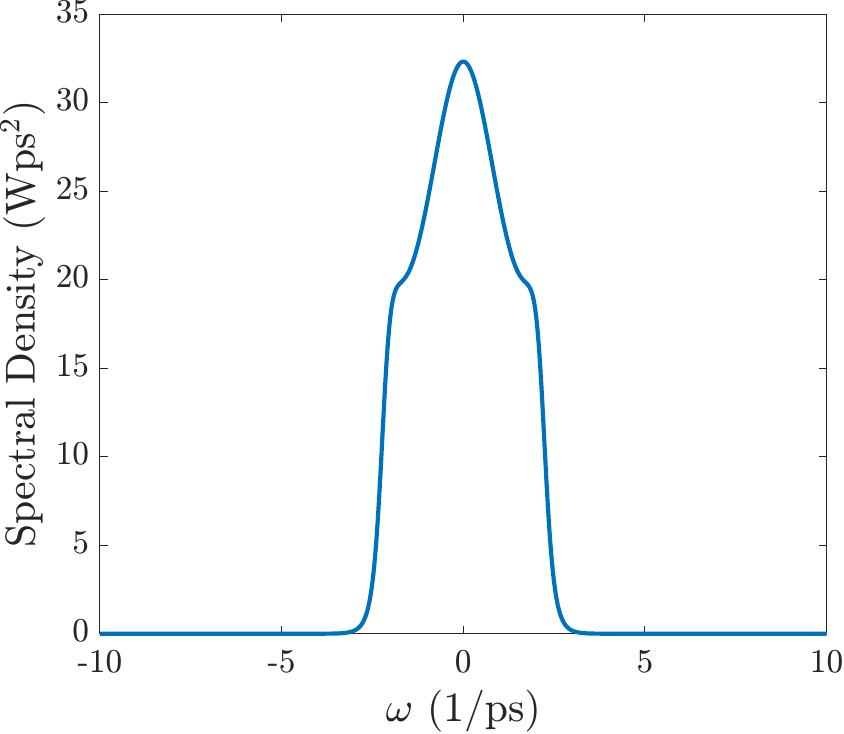}
        \label{LAS-spect}}
\caption{\label{LAS} Power profile and chirp \subref{LAS-PowerChirp} and spectral density \subref{LAS-spect} of the LAS for $g_0=1.465$ m$^{-1}$, $T_2=100$ fs and $P_\text{sat}=80$ W. The dotted line corresponds to  $P_p\,\text{sech}\,^2(t)$, with $P_p$ the peak power.}
\end{figure}

The power profile given by $|W(t)|^2$, the chirp defined by $-d\phi/dt$ ($\phi$ is the phase of $W(t)$) and the spectrum profiles are shown in graphs of Figs.~\ref{LAS}, \ref{MAS} and \ref{HAS}. The LAS power profile has wider tails than a sech$^2$, linear chirp at the pulse peak and constant chirp at the tails and its spectral profile shows one single peak at the carrier frequency (Fig.~\ref{LAS}). As mentioned in the introduction, all the solutions of this type were found to be unstable, having one unstable eigenvalue that is purely imaginary.

The solutions named MAS have different characteristics depending on the value of $T_2$. For lower $T_2$, the power profile is close to a sech$^2$ ($P_p\,\text{sech}^2(t)$, with $P_p$ being the peak power) on top of a pedestal. The chirp is also linear at the pulse peak position and constant at the tails (similar to the LAS chirp profile) and the spectrum exhibits two distinguished peaks as reported before \cite{quian22} (Figs.~\ref{MAS-PowerChirp} and \ref{MAS-spect}). For higher $T_2$, the power profile is close to $P_p\,\text{sech}^2(t/\zeta)$, with $\zeta>1$, thus, it is wider than the $P_p\,\text{sech}^2(t)$ that adjusts to the central part of the MAS solutions for lower $T_2$. The chirp is also similar to the ones referred before for LAS and MAS with low $T_2$ but showing less structure, as may be observed in Fig.~\ref{MAS2-PowerChirp}, and the spectrum is single peaked (Fig.~\ref{MAS2-spect}). 
These characteristics were already reported by us in \cite {malheiro23} and are analogous to those previously found in {\cite{quian22,wang24,wu24}.

\begin{figure}[hbt!]
    \subfigure[]{
        \centering
        \includegraphics[height=0.25\textheight]{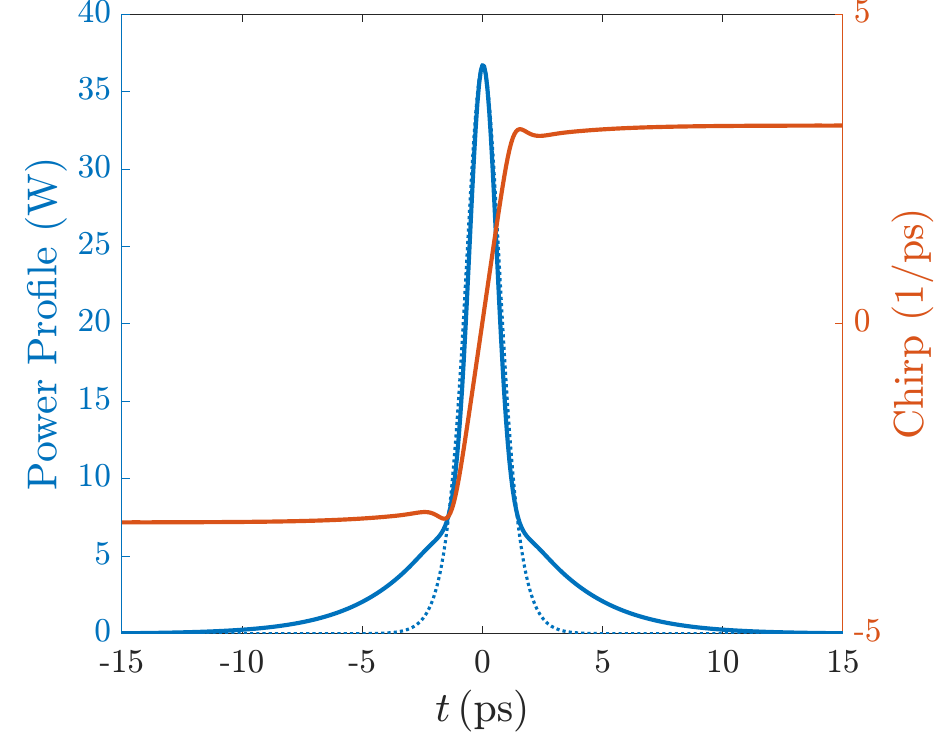}
        \label{MAS-PowerChirp}}
    \subfigure[]{
        \centering
        \includegraphics[height=0.25\textheight]{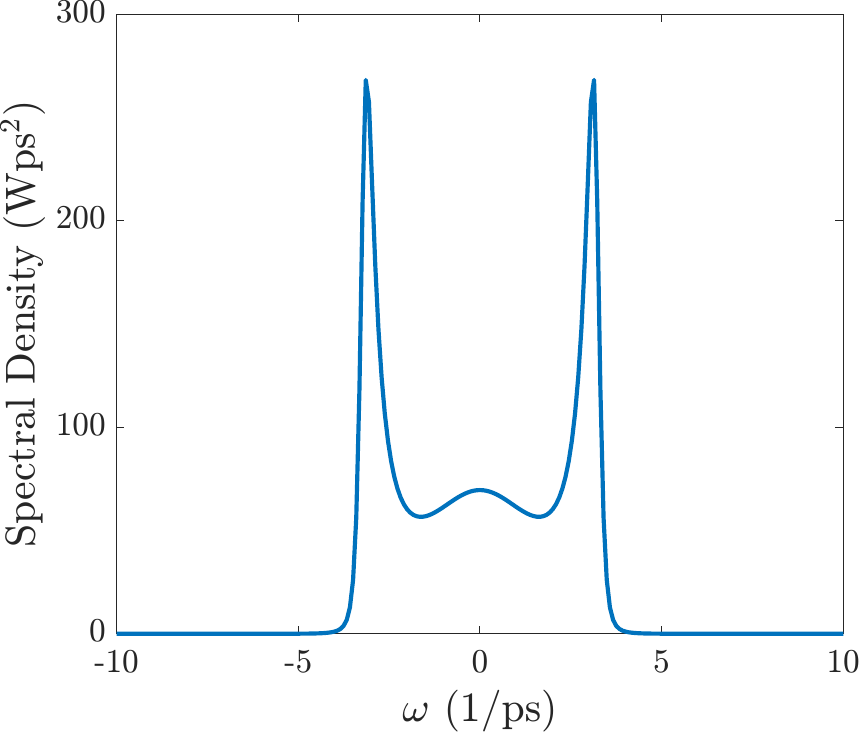}
        \label{MAS-spect}}
        
    \subfigure[]{
        \centering
        \includegraphics[height=0.25\textheight]{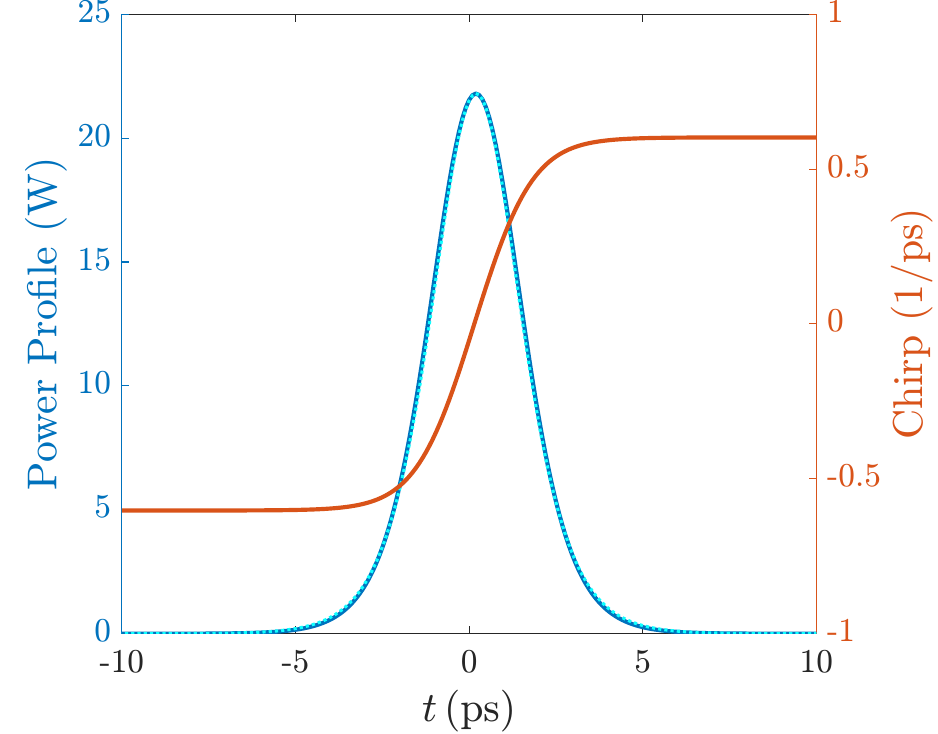}
        \label{MAS2-PowerChirp}}
    \subfigure[]{
        \centering
        \includegraphics[height=0.25\textheight]{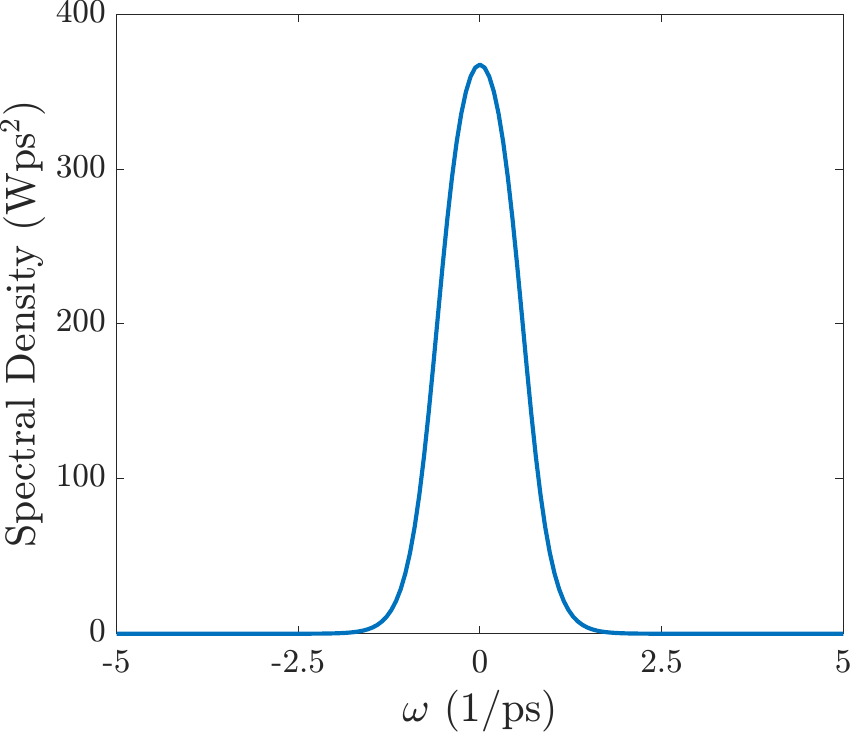}
        \label{MAS2-spect}}
\caption{\label{MAS} Power profile and chirp and spectral density  of the MAS for $g_0=1.465$ m$^{-1}$, $T_2=100$ fs and $P_\text{sat}=80$ W \subref{MAS-PowerChirp} and \subref{MAS-spect} and for $g_0=1.485$ m$^{-1}$, $T_2=550$ fs and $P_\text{sat}=80$ W \subref{MAS2-PowerChirp} and \subref{MAS2-spect}. The dotted lines correspond to  $P_p\,\text{sech}^2(t)$ in \subref{MAS-PowerChirp} and to $P_p\,\text{sech}^2(0.58 t)$ in \subref{MAS2-PowerChirp}.}
\end{figure}

Finally, the HAS solutions have almost triangular power profiles, exhibiting an extra sharp peak at the top, a chirp profile close to a single step (Fig.~\ref{HAS-PowerChirp}) and the spectrum has two pronounced peaks (see Fig.~\ref{HAS-spect}). These characteristics are similar to those of the triangular solution previously reported in \cite{runge20OL} for a different mode-locked laser model and in \cite{wang24} for the CGLE.
We found that the tails of the triangular amplitude profile can be well approximated by a function of the type 
\begin{equation}
   |W(t)| = a\exp\left[-\left(\frac{|t|}{c}\right)^b\right],
    \label{eq:HAS_fit}
\end{equation}
where $a$, $b$ and $c$ are real constants. Such a fit is represented in Fig.~\ref{HAS-PowerChirp} and the dependence of $a$, $b$ and $c$ with $T_2$ for several $g_0$ is illustrated in Fig.~\ref{fig:HAS_fit}. We found that the value of $b$ remains similar, around 1.5, when $T_2$ and $g_0$ are changed. The $a$ value, which is associated with the amplitude, and the $c$ value, associated with the width, both decrease with $T_2$ and increase with $g_0$. The full width at half maximum (FWHM, in all remaining text named $\tau$) of the power profile $|W(t)|^2$ with $W(t)$ given by Eq.~(\ref{eq:HAS_fit}) is easily obtained as $\tau=2c(\log 2/2)^{1/b}$ and the energy $E=\int|W(t)|^2dt$ is given by $E=a^2(\log 2)^{-1/b}\Gamma(1+1/b)\tau$, with $\Gamma$ representing the Gamma function. This expression for the energy includes both $a$ and $b$. Since $b$ is almost constant, we may only understand the relation between $a^2$ and width. For that purpose, we graphed both quantities as shown in figure \ref{a2_versus_c}, which reveals that they are almost linear. Thus, we estimate that the energy of the HAS scales approximately with $\tau^2$.

\begin{figure}[hbt!]
    \subfigure[]{
        \centering\includegraphics[height=0.25\textheight]{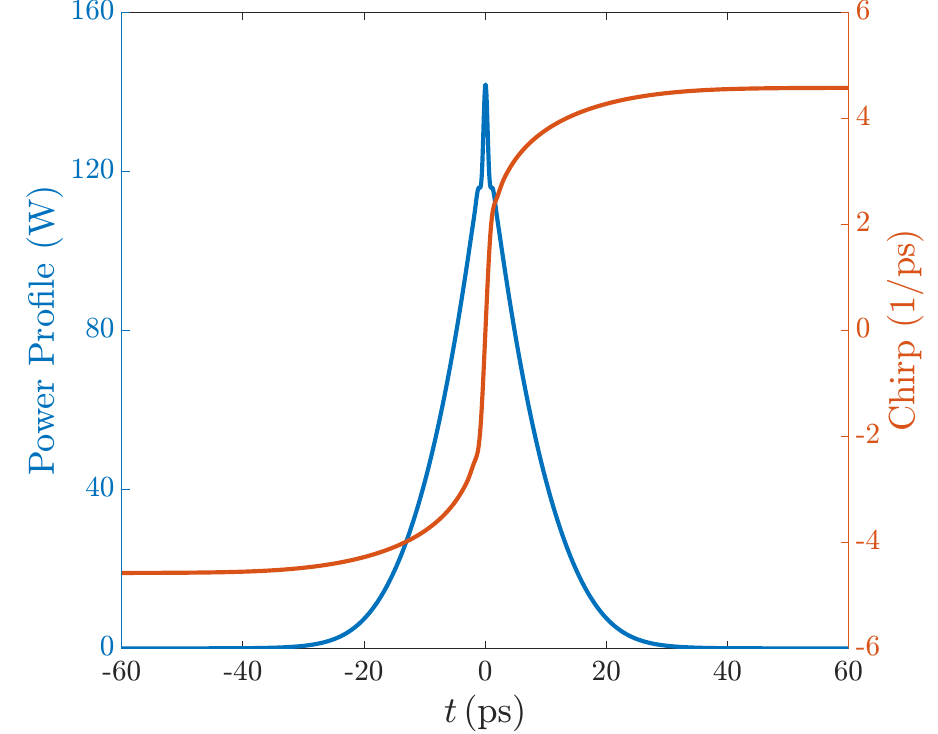}
        \label{HAS-PowerChirp}}
    \subfigure[]{
        \centering\includegraphics[height=0.25\textheight]{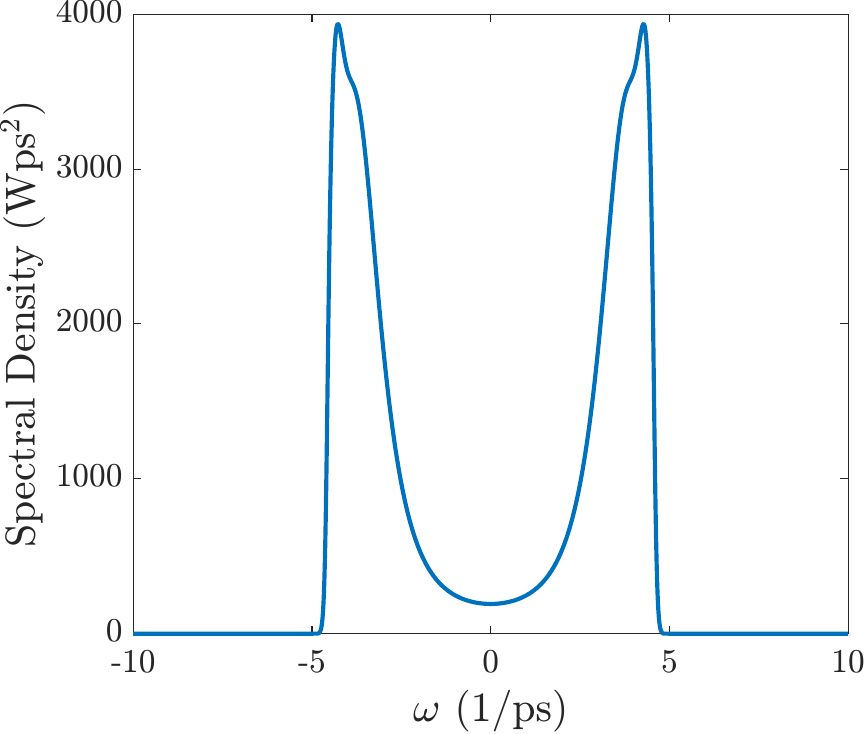}
        \label{HAS-spect}}
\caption{\label{HAS} Power profile \subref{HAS-PowerChirp} and spectral density \subref{HAS-spect} of the HAS for $g_0=1.465$ m$^{-1}$, $T_2=100$ fs and $P_\text{sat}=80$ W. The dashed line represent a fit to the amplitude profile of the type of Eq. \eqref{eq:HAS_fit}, with $a = 10.99$, $b = 1.43$ and $c = 15.75$.}
\end{figure}

\begin{figure}[hbt!]
\centering
    \subfigure[]{\centering
    \includegraphics[height=0.25\textheight]{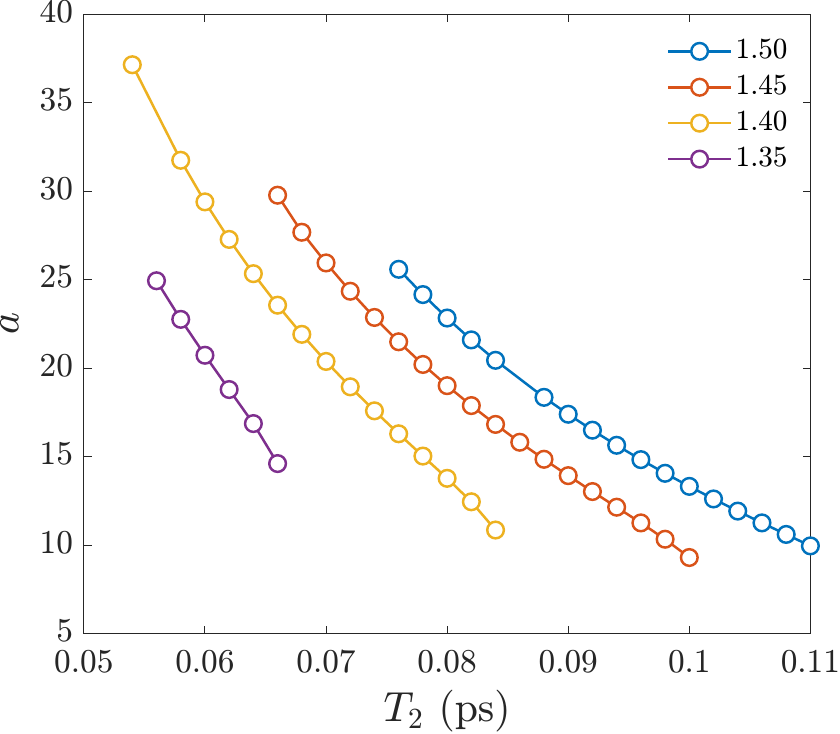} \label{fig:HAS_fit_a}}
     \subfigure[]{\centering\includegraphics[height=0.25\textheight]{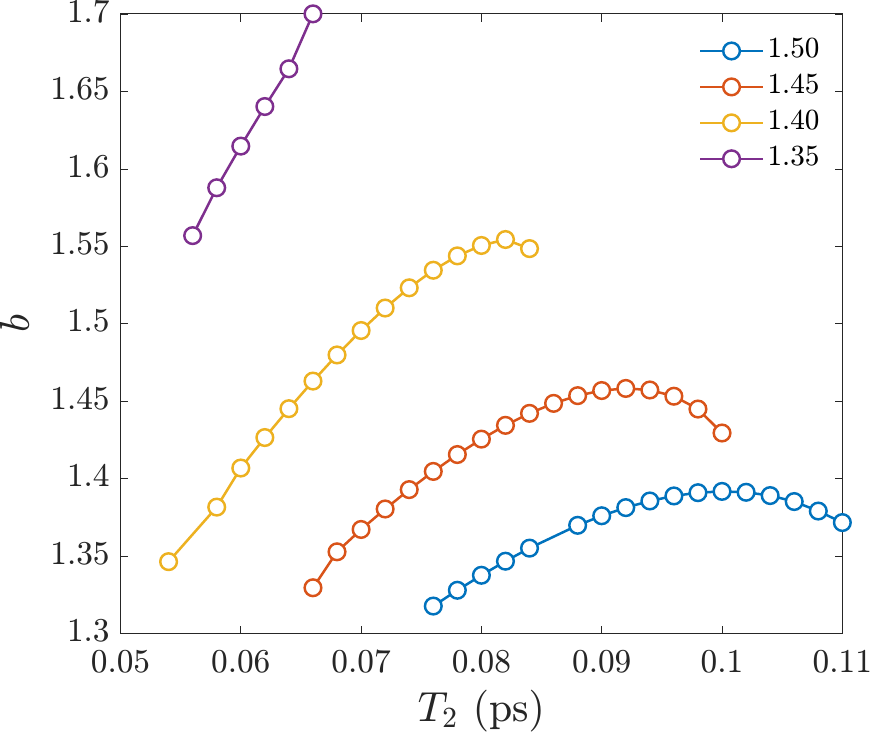}\label{fig:HAS_fit_b}}
      \subfigure[]{\centering\includegraphics[height=0.25\textheight]{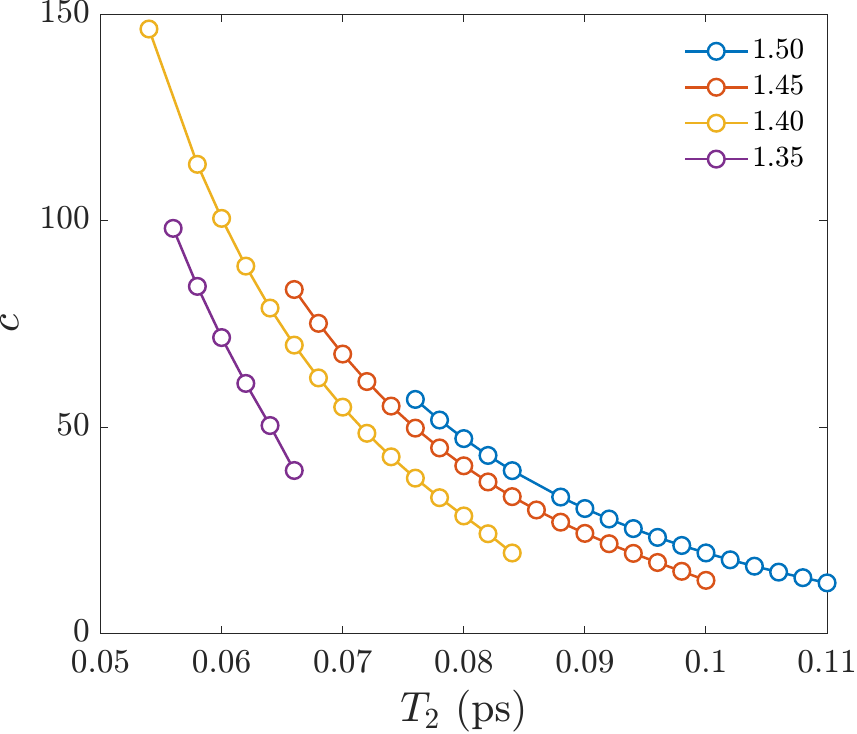}\label{fig:HAS_fit_ac}}
      \subfigure[]{\centering\includegraphics[height=0.25\textheight]{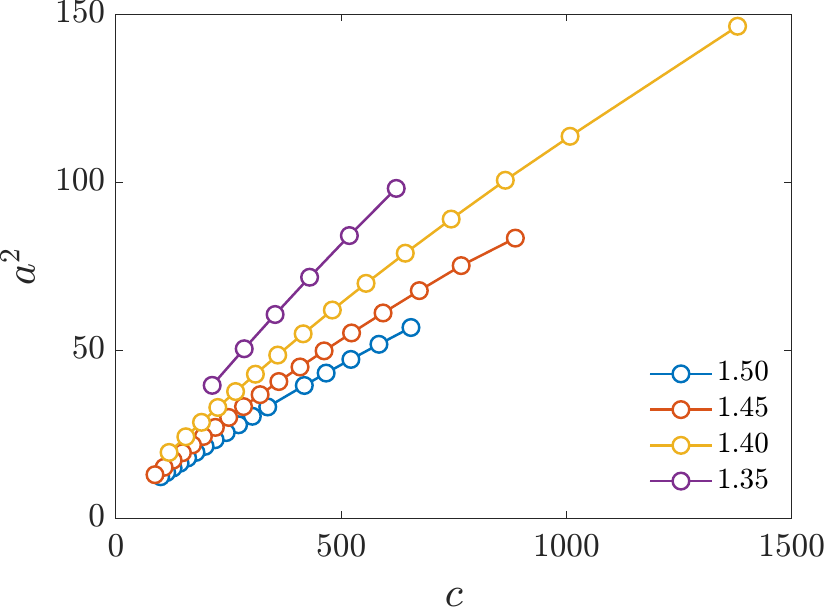}\label{a2_versus_c}}
    \caption{Values of $a$, $b$ and $c$ obtained from fitting Eq.~\eqref{eq:HAS_fit} to HAS amplitude profiles as function of $T_2$ and $a^2$ versus $c$ for several values of $g_0$ (m$^{-1}$) as indicated in the legends. The different ranges of $T_2$ here presented are related to solution existence and stability (see Section C, Fig. \ref{HAS_Ew_1})}
    \label{fig:HAS_fit}
\end{figure}

As was already commented in the introduction, the cubic-quintic CGLE has been widely used to model mode-locked lasers. We may reach the cubic-quintic CGLE by expanding the term of the saturable absorber up to the fifth order in $W$, but this should only be valid for small $|W|^2/\bar{P}_\text{sat}$. The examples shown in Figures \ref{LAS}, \ref{MAS} and \ref{HAS} have peak power values of 7.8~W (LAS), 36 and 22~W (MAS) and 140~W (HAS) for a $\bar{P}_\text{sat}$ value around 18~W. Thus, among these examples, only the LAS could be reasonable obtained using the cubic-quintic CGLE. Nevertheless, we found LAS with peak values as large as 20~W (Fig.~\ref{bifurc}), MAS peaking at 50~W, for $g_0=1.501~\mathrm{m}^{-1}$ and $T_2=124$~fs, and HAS peaking at 1600~W, for $g_0=1.37~\mathrm{m}^{-1}$ and $T_2=51$~fs, which are results that invalidate the expansion of the saturable absorber.

\subsection{Energy Flow} \label{sec:energy_flow}
In order to understand the energy exchange dynamics to the exterior and within the pulse, we use the following continuity relation for equation (\ref{PDE2}) \cite{dissipative}
\begin{equation}
    \frac{\partial s}{\partial Z} + \frac{\partial j}{\partial T}=P.
    \end{equation}
In the above equation, $s=|q|^2$ is the density of power (power per unit time), which is constant along $Z$ for stationary solitons. $j$ is a flux of density of power, i.e., $j=su$ with $u$ being a velocity (derivative of $T$ in order to $Z$) indicating the direction, in $T$, of the movement of power density as $Z$ is varied, given by
\begin{equation}
j=i\frac{D_4}{24}\left(qq_{TTT}^*-q_{TTT}q^*+q_{TT}q_T^*-q_Tq_{TT}^*\right).\end{equation}
The flux of power $j$ is associated with the conservative terms, in this case, the dispersion term, and reveal the direction of the flow of power within the pulse, in parts at which $j$ is positive the flux is in the direction of positive $T$ and if negative the direction of the flux is towards negative $T$.
$P$ is the density of power per distance $Z$ that enters or leaves the system from or to the exterior. given by
\begin{equation}
P=2\alpha|q|^2+(|q|^2)_{TT}-2|q_T|^2-2(1+\alpha)\frac{|q|^2}{1+\rho|q|^2}.\end{equation}
$P$ is related with the dissipative terms, it is positive in parts of the pulse at which energy is absorbed from the medium and negative in parts at which the pulse dissipates energy. 

The graphs of Fig.~\ref{flux} show $j$ and $P$ for all the solutions of Figs.~\ref{LAS}-\ref{HAS}. 
For all solutions, the energy enters through the center of the pulse, in regions of positive $P$ that are identified in the graphs in gray, and then is routed to the tails where it is dissipated. This behaviour was expected since the saturable absorber is saturated in the center, allowing for nonlinear gain to manifest (thus, energy is entering the pulse at the center), and is unsaturated at the tails, where losses will occur.

\begin{figure}[hbt!]
 \subfigure[]{\centering
        \includegraphics[width=0.48\textwidth]{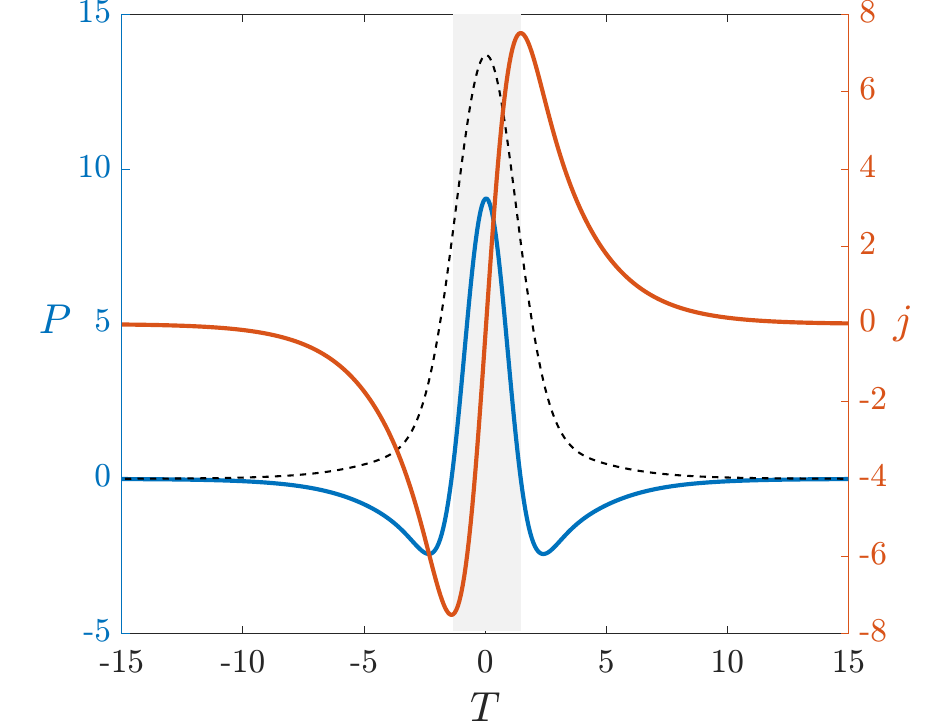}  \label{LAS_flux}}  
  \subfigure[]{\centering
        \includegraphics[width=0.48\textwidth]{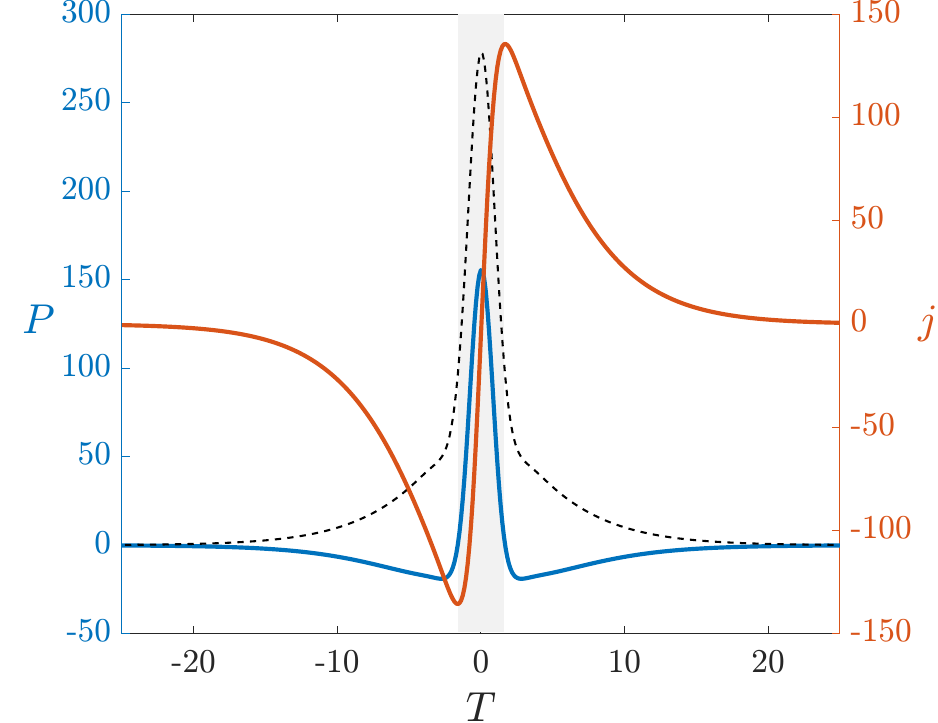}  \label{MAS_flux}}  
        
        \subfigure[]{\centering
        \includegraphics[width=0.48\textwidth]{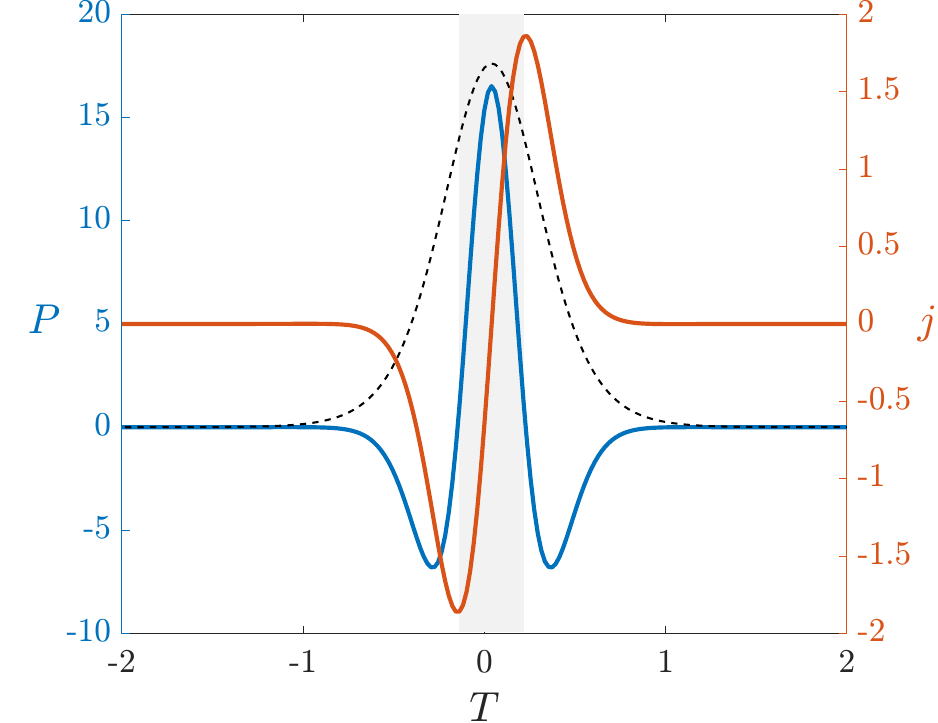}  \label{MAS2_flux}}  
         \subfigure[]{\centering
        \includegraphics[width=0.48\textwidth]{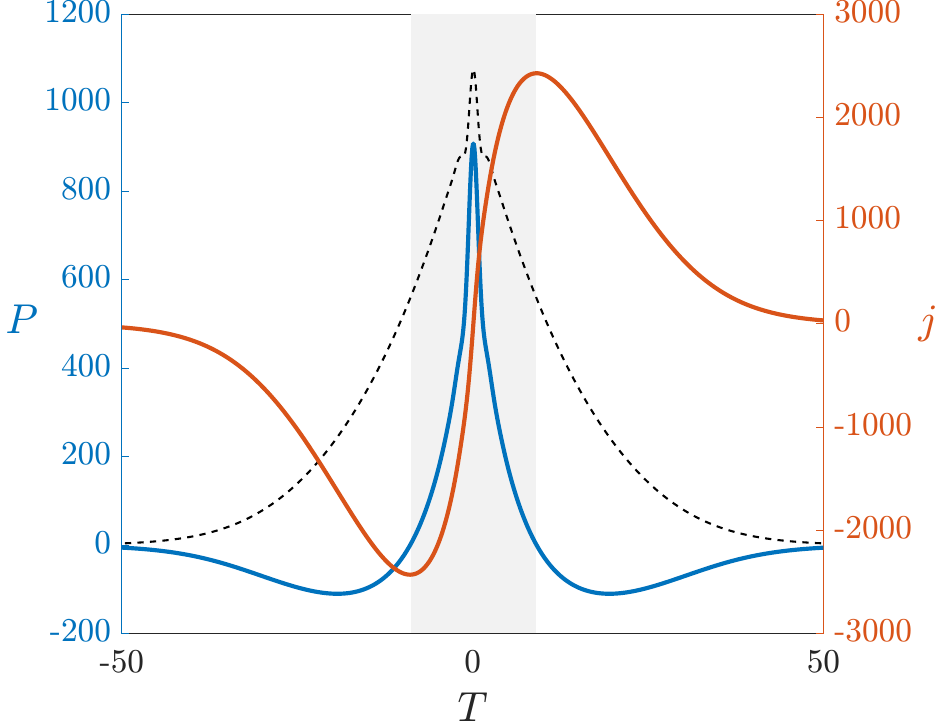}  \label{HAS_flux}}    
\caption{\label{flux}Density of energy generation (blue line) and internal flux of energy (red line) both graphed versus temporal position of the LAS for $g_0=1.465$ m$^{-1}$, $T_2=100$ fs \subref{LAS_flux}, MAS for $g_0=1.465$ m$^{-1}$, $T_2=100$ fs \subref{MAS_flux}, MAS for $g_0=1.485$ m$^{-1}$, $T_2=550$ fs \subref{MAS2_flux} and HAS for $g_0=1.465$ m$^{-1}$, $T_2=100$ fs \subref{HAS_flux}. All the quantities are dimensionless. The dashed lines are the corresponding power profiles with the correct time scale (dimensionless) but with arbitrary units in amplitude, to serve  as a reference. The gray dashed zones identify the temporal range at which $P>0$.} 
\end{figure}
\newpage
\subsection{Existence, stability and bistability} \label{sec:stability}
The three main solutions are related by a bifurcation kind of dynamics that is shown in Fig.~\ref{bifurc} for the particular value of $g_0=1.465~\text{m}^{-1}$. The graph shows the peak power as $T_2$ is swept for the three types of solutions. The lower branch corresponds to LAS which is connected to the MAS branch at a higher threshold $T_2$ value (approximately 200~fs in the case shown in the graph). This connection reveals characteristics of a saddle-node bifurcation, at which a stable and an unstable solutions collide at some parameter value and no other solutions exist beyond the same parameter value.  There are stable and unstable MAS solutions, stable for higher $T_2$ but unstable below a particular value of $T_2$ (approximately 100~fs in the case shown in the graph). The HAS bifurcates in a complicated way from the branch of MAS.
LAS, MAS and HAS solutions can coexist, either all three at once, or in pairs, as well as with the other more restricted branches of solutions represented in Fig.~\ref{bifurc}. The solutions in these other branches are all unstable and were found by integrating the ODE for different values of $T_2$, starting the NCG method with a previous solution, and modifying $T_2$ in different steps and in opposite directions. The existence of several consecutive unstable branches had already been observed on the CGLE, for instance, in \cite{sotocrespo01pla} for anomalous 2OD. The profiles of the pulses of those unstable branches are shown in Fig.~\ref{puls_bifurc} and all of them exhibit the sharp pulse on top of a larger part that is characteristic of the HAS solutions. Solutions identified as C and D are actually quite similar to the HAS solutions. In the other solutions represented, the larger part of the pulse is seen to be composed of two different regions. This characteristic is more pronounced in pulse F.

MAS and HAS are both stable in a region of parameters, as shown below, producing bistability behaviors. One of these behaviors is hysteresis as may be observed in the results shown on Fig.~\ref{hysteresis}. We solved Eq.~(\ref{PDE2}) using inputs from previous simulations for close values of $T_2$. The results labeled as 'Increasing $T_2$' correspond to the output solutions obtained with inputs of lower $T_2$ and the ones labeled as 'Decreasing $T_2$' correspond to the output solutions obtained with inputs of higher $T_2$. In fact, in the coexistence region the results obtained with different inputs do not coincide. In the direction of decreasing $T_2$, the jump between the two type of solutions is abrupt. However, in the other direction, there is a range of $T_2$ (105 to 109 fs) for which the simulation did not produce a stationary pulse but instead a pulse that evolved in $Z$ with peak and form variations, showing some similarities with the breathing behaviour that was already reported for dissipative quartic solitons but for negative 4OD \cite{zhang22}. We anticipate that this oscillatory behavior can be explained by the existence of different branches of unstable solutions in this region. Figure \ref{breather} shows two kinds of evolution, one converging to a stationary soliton and the other showing the non-stationary evolution referred for the range of $T_2$ between 105 to 109 fs.

\begin{figure}[hbt!]
\centering
        \subfigure[]{\label{bifurc}\centering
        \includegraphics[width=0.48\textwidth]{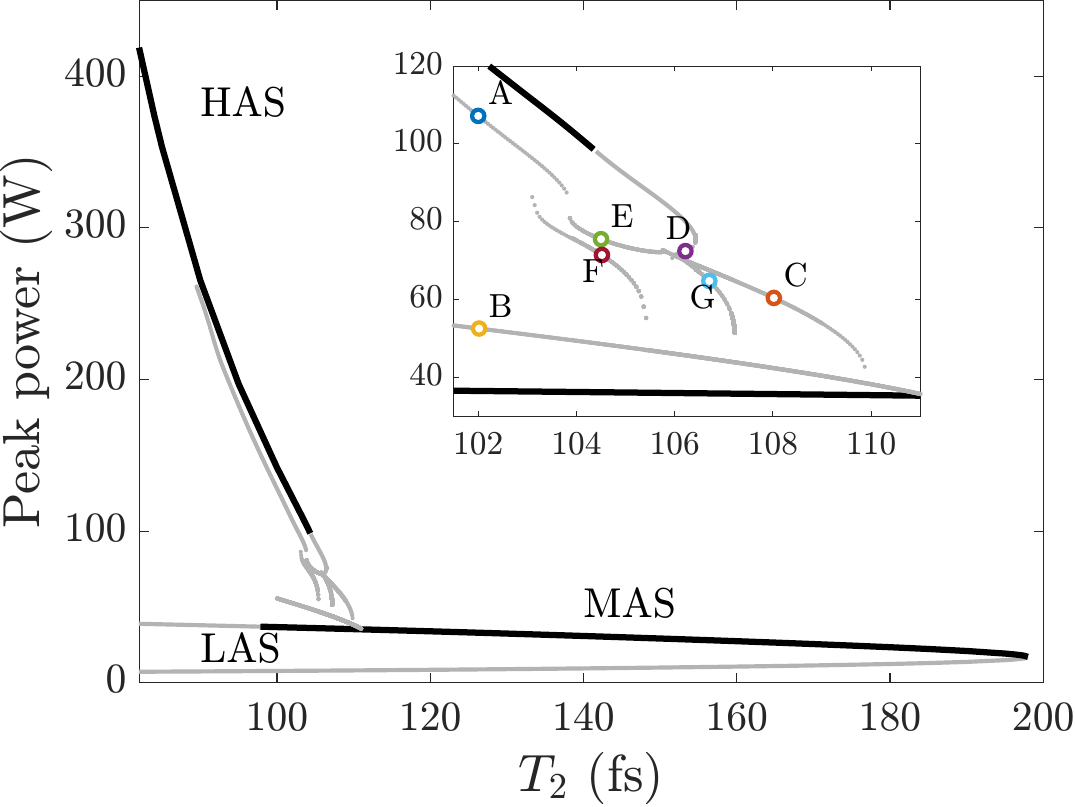}   }  
          \subfigure[]{\label{puls_bifurc}\centering
        \includegraphics[width=0.48\textwidth]{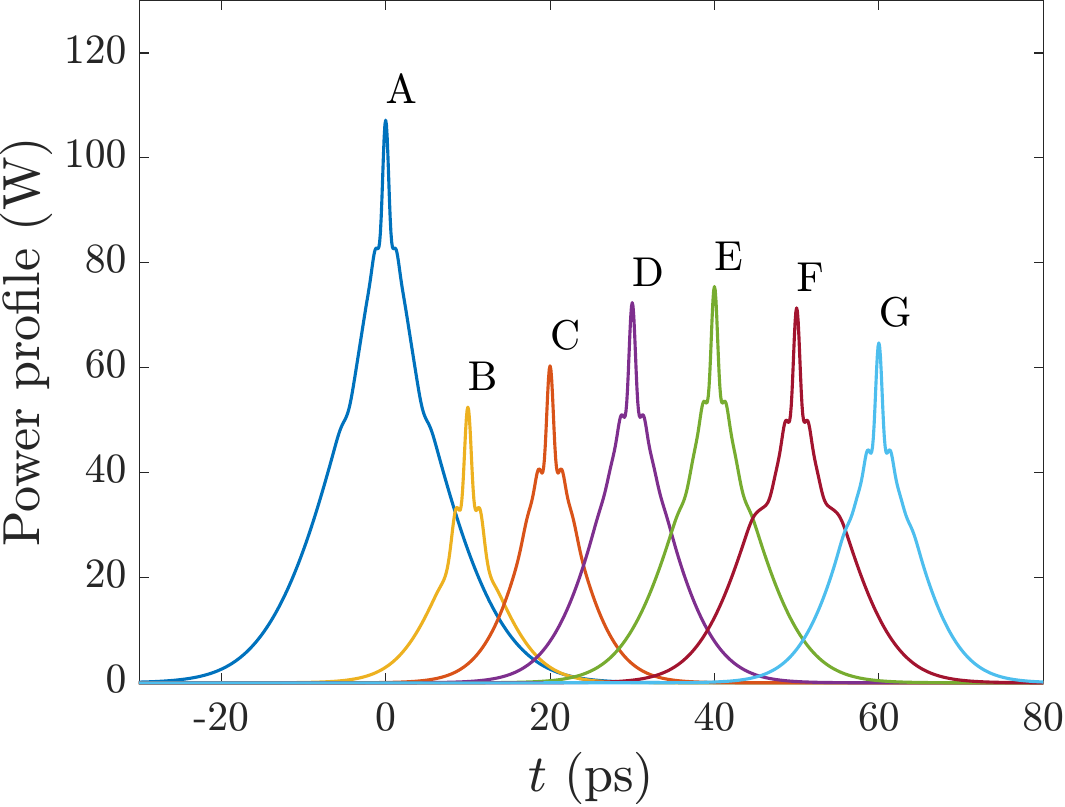}   } 
        \subfigure[]{\centering\label{hysteresis}
    
    \includegraphics[width=0.48\textwidth]{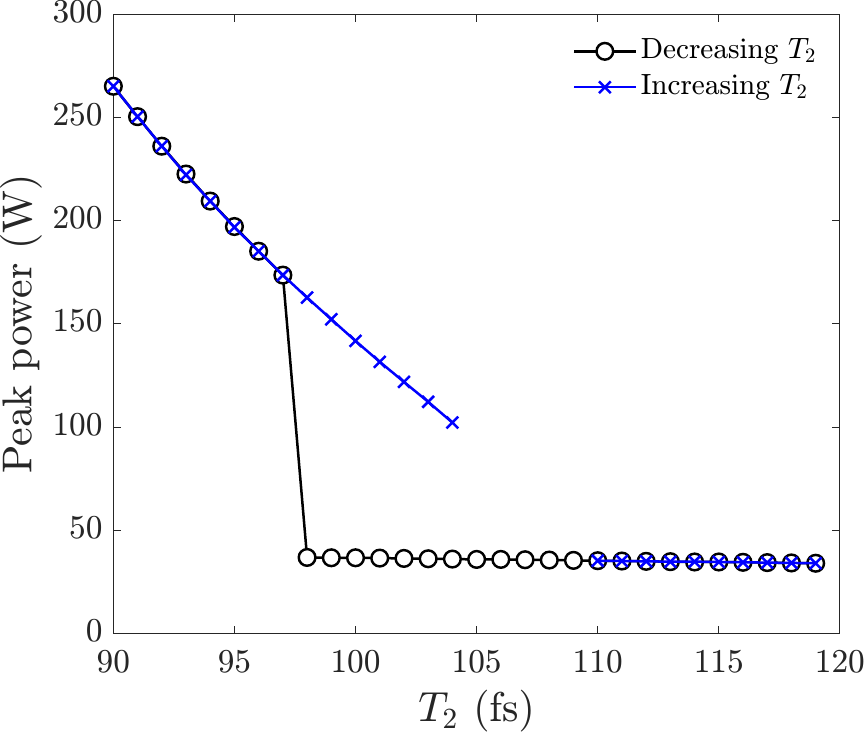}    }
           
\caption{\subref{bifurc} Bifurcation diagram showing the peak power versus $T_2$ with several branches of solutions (including LAS, MAS and HAS). Stable and unstable branches are represented by black and gray lines, respectively. 
\subref{puls_bifurc} Power profiles for the solutions identified in \subref{bifurc}.
\subref{hysteresis} Hysteresis shown by the peak value of the output pulses obtained by solving Eq.~(\ref{PDE2}) with inputs from higher $T_2$ (Decreasing $T_2$) and inputs from lower $T_2$ (Increasing $T_2$). 
The results shown correspond to $P_\text{sat}=80$~W and $g_0=1.465$~m$^{-1}$.}
\end{figure}

\begin{figure}[hbt!]
\centering
        \subfigure[]{\label{100}\centering
        \includegraphics[width=0.48\textwidth]{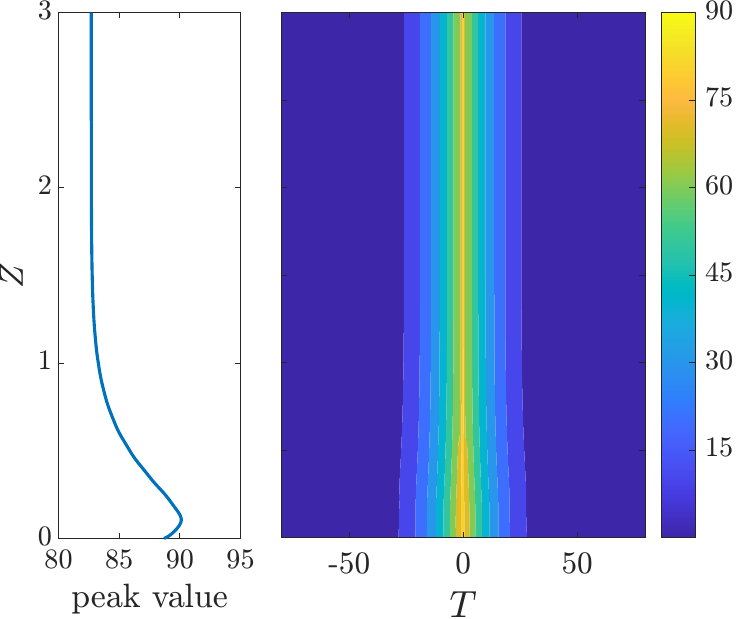}}
        \subfigure[]{\label{107}\centering
        \includegraphics[width=0.48\textwidth]{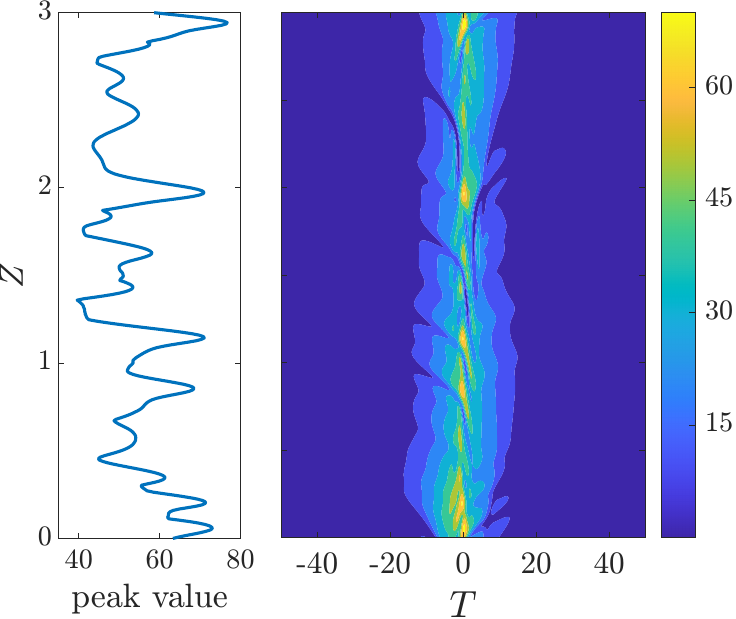}}
\caption{\label{breather}Evolution of peak power and of the full power profile (illustrated by a contour plot) in the case of stationary evolution \subref{100} for $g_0=1.465$~m$^{-1}$ and $T_2=100$~fs and non-stationay evolution \subref{107} for $g_0=1.465$~m$^{-1}$ and $T_2=107$~fs, both in adimensional variables. The input was the output of a previous simulation for the same $g_0$ and a lower $T_2$. }
\end{figure}

Restricting our attention to possible stable solutions, namely, to MAS and HAS solutions, we scanned the $(g_0,T_2)$ space to find regions at which those solutions exist and are stable. The regions are those in Figs.~\ref{MAS_Ew_1} and \ref{HAS_Ew_1} where energy and temporal width are also shown. For increased clarity, the boundaries of the graphs should be explained. The right boundary is at $g_0=1.502~\text{m}^{-1}$ very close to the limit 1.504 above which the linear loss would become linear gain and the background would become unstable. Above that boundary, even if solitons exist, they would not survive in the unstable background. The upper boundary of MAS, for $T_2$ higher than 880 fs and $g_0>1.48$, is not an actual boundary but only a limit at which the scanning of the $(g_0,T_2)$ space was stopped. 

\begin{figure}[hbt!]
    \subfigure[]{
        \centering\includegraphics[width =0.48\textwidth, angle=0]{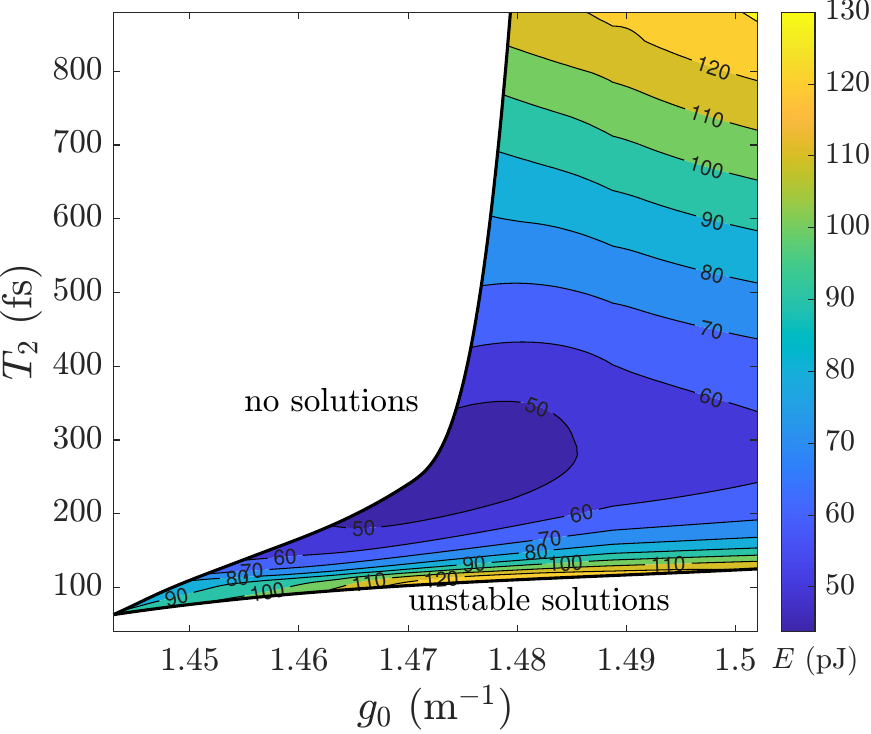}
        \label{MAS_energy}}
    \subfigure[]{
        \centering\includegraphics[width =0.48\textwidth, angle = 0]{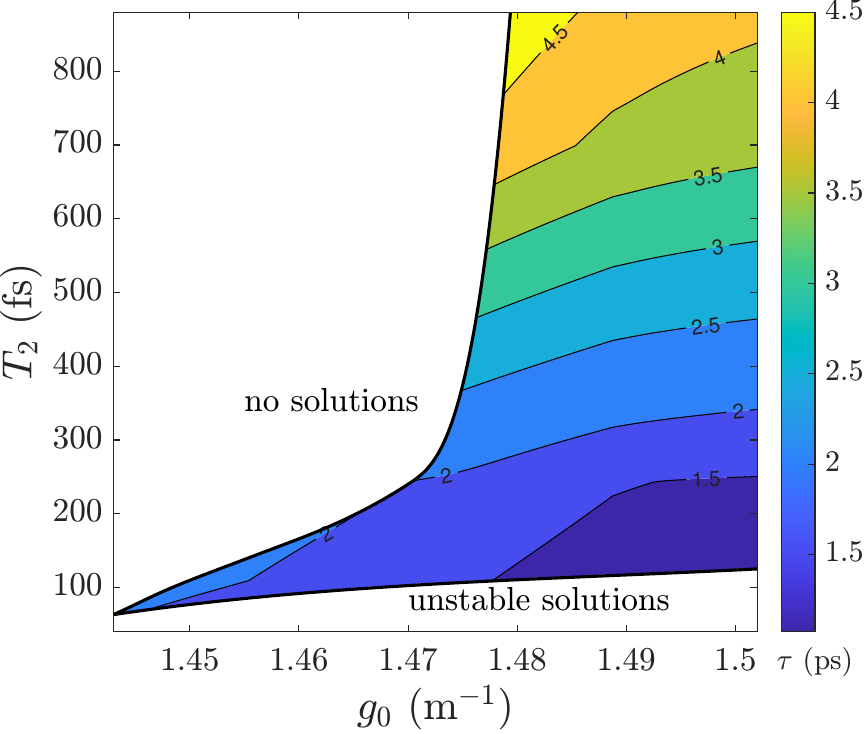}
        \label{MAS_width}}
\caption{\label{MAS_Ew_1} Contour plots of the energy in pJ \subref{MAS_energy} and pulse width in ps \subref{MAS_width} for MAS.}
\end{figure}

Concerning the region of existence and stability of MAS, it was found to be composed by two distinguishable bands. The first, exists for a $g_0$ range between 1.443 and 1.475~m$^{-1}$, being very limited to $T_2$ values around 100 fs. The second band exists in a much broader $T_2$ region, with $T_2 \gtrsim 100$, but is much more limited in $g_0$, ranging from 1.475 to 1.5~m$^{-1}$.
Note that the results for $\beta_4>0$ presented in \cite{malheiro23} were for MAS on the first referred thin band of existence. The lower boundary of the MAS region corresponds to a transition to unstable solutions at which two complex conjugated stability eigenvalues cross the real axis from the stable half plane to the unstable half plane. On the other hand, above the upper boundary, no soliton solutions were found. Figs.~\ref{MAS_energy} and \ref{MAS_width} also have contour lines that show that, for lower values of $T_2$ the most energetic pulses are also the shortest ones. As noted in \cite{malheiro23}, the energy increase with $g_0$ at a fixed $T_2$ is explained by the fact that the former parameter is directly related to energy gain. The energy dependence with $T_2$ however is not as straightforward to interpret, as it can be non-monotonous in some cases (assuming constant $g_0$). In terms of the width, it can be seen that, for MAS, an increase in $T_2$ tends to lead to wider pulses. $T_2$ represents the inverse linewidth of the parabolic gain and, therefore, a higher $T_2$ could justify larger pulse widths. However, as was shown in Fig. \ref{fig:HAS_fit}, for HAS solutions the width actually decreases with $T_2$, showing that there are indeed exceptions to this trend.

\begin{figure}[hbt!]
    \subfigure[]{
        \centering\includegraphics[width =0.48\textwidth]{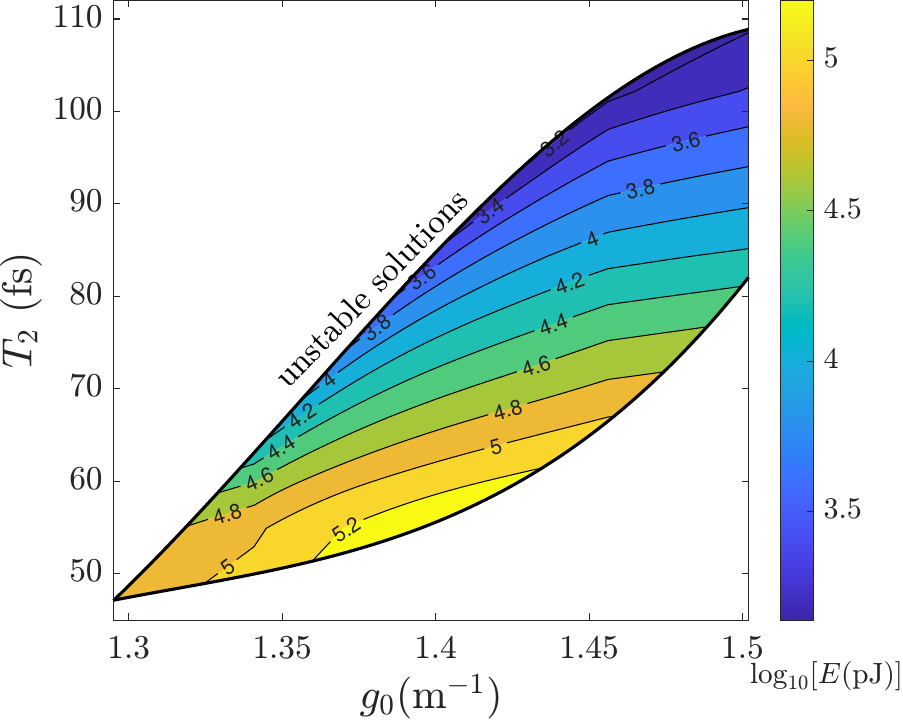}
        \label{HAS_energy}}
    \subfigure[]{
        \centering\includegraphics[width =0.48\textwidth]{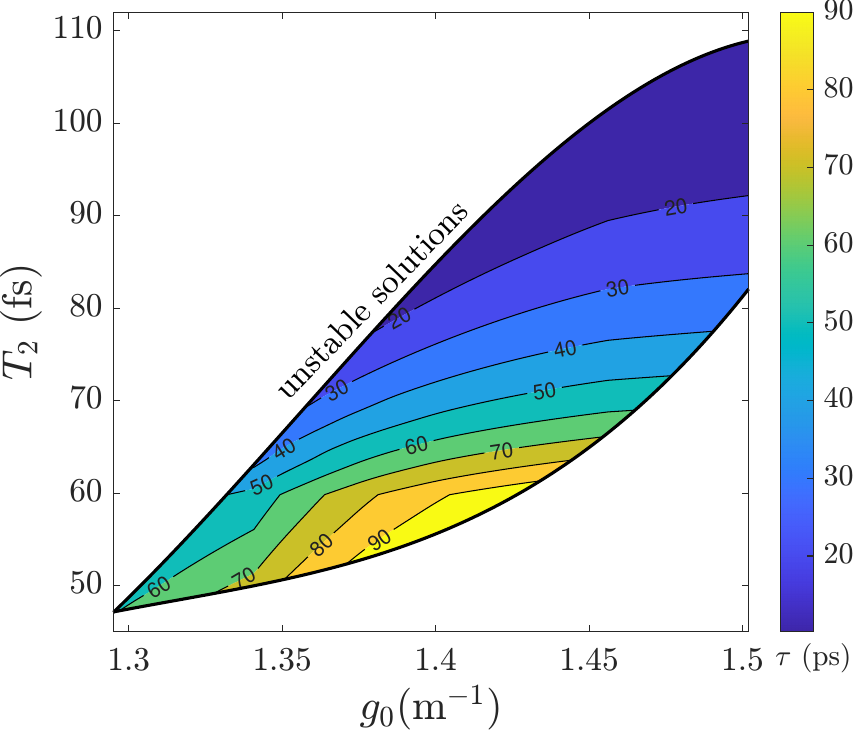}
        \label{HAS_width}}
\caption{\label{HAS_Ew_1} Contour plots of base 10 logarithm of the energy in pJ \subref{HAS_energy} and pulse width in ps \subref{HAS_width} for the HAS.}
\end{figure}

Regarding HAS, the region of existence is larger in terms of $g_0$, from $1.295~\text{m}^{-1}$ to $1.50~\text{m}^{-1}$, but smaller in terms of $T_2$ which, in the largest region, may only be varied from 60 fs to 100 fs (see Fig.~\ref{HAS_Ew_1}).  As the upper border is crossed the solutions continue to exist, at least in the region in close proximity with the border, but are unstable, having a pair of complex unstable eigenvalues. Below the lower border, it was not possible to check for existence or instability of solutions since the NCG method did not converge efficiently. Thus, that border was found by observation of pulse propagation as given by the direct integration of Eq.~(\ref{PDE2}), which revealed destruction of the pulses starting in their slide slopes. The contour lines of Fig.~\ref{HAS_Ew_1} show that, contrary to what is observed for the MAS, the more energetic pulses correspond to wider pulses. This topic will be further explored below. Moreover, to obtain highly energetic pulses, $T_2$ should be lowered and $g_0$ increased. These actual results are in agreement with the energy of the approximate profile in Eq.~(\ref{eq:HAS_fit}) and Fig.~\ref{fig:HAS_fit_a} since we have shown that the energy scales with $a^2$ and $a$ is higher for lower $T_2$ and higher $g_0$. 

As we referred above, the overlap of regions of stable MAS and HAS is not null, i.e., there are parameter values for which both MAS and HAS solutions exist and are stable. This parameter region of bistability is better observed in Fig.~\ref{MAS-HAS}, showing it to be small when compared with regions of existence of each separate type of soliton. 

Our MAS and HAS should correspond to the dissipative pure quartic solitons (DPQSs) and quartic self-similar pulses (QSSPs), respectively, both referred in \cite{wang24}. In fact, the solutions here reported are for an equation that is not the CGLE, however, if we expand the saturable absorber term, we should reach the CGLE. Increasing $T_2$ will have the effect of decreasing $D_4$ and maintaining all the other CGLE parameters fixed. Our results show that increasing $T_2$ may cause the HAS to cease to exist giving rise to MAS, which is consistent with results in \cite{wang24} for decreasing $D_4$. The increase of $g_0$ will produce both higher nonlinear gain but also lower $D_4$, thus the comparison of our results with results in \cite{wang24} is not so straightforward.

\begin{figure}[hbt!]
        \centering
        \includegraphics[width=0.5\textwidth]{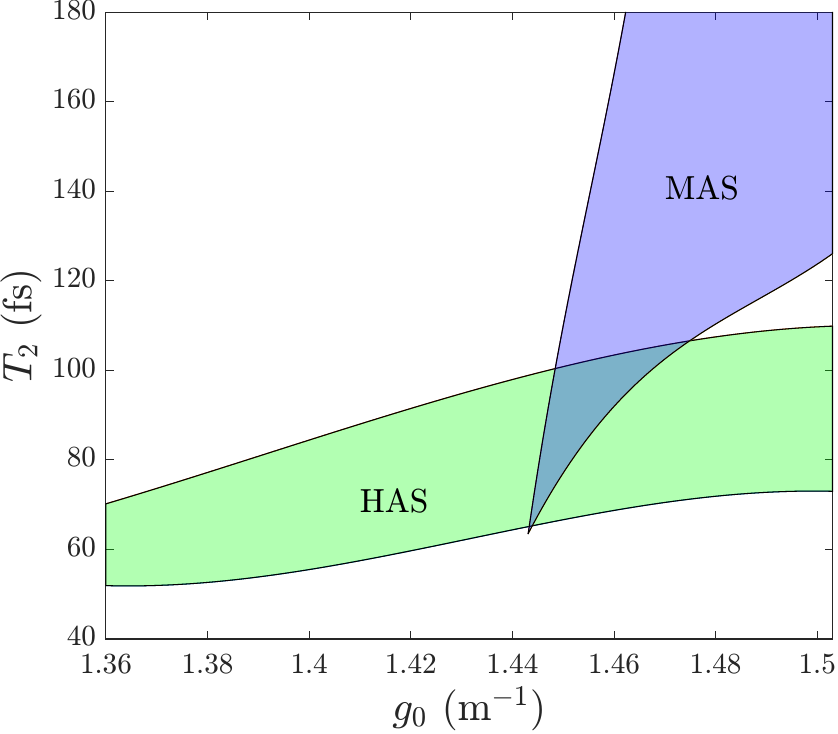}      
\caption{Graph showing the overlap of stable MAS (blue) and HAS (green) parameter space regions, considering $P_\text{sat}=80$ W.}\label{MAS-HAS}
\end{figure}

In parameter regions where only HAS solutions are stable, it is also possible to observe the evolution from unstable LAS and/or MAS solutions into a stable HAS, provided that the first two solutions exist in such a parameter region. For example, for $g_0 = 1.47~\text{m}^{-1}$ and $T_2 = 100~\text{fs}$, a LAS solution was obtained from solving Eq. \eqref{ODE} through the NCG method. When such a solution is used as input for Eq. \eqref{PDE2}, it vanishes during propagation. If it is slightly perturbed, however, it will transition to an MAS solution, which is also unstable, propagating a certain distance with rising-amplitude oscillations in the peak power, until it abruptly transitions into the stable HAS solution. This behavior is illustrated in Fig.~\ref{fig:solution_transition}, with Fig.~\ref{fig:peak_power_transition} showing the evolution of the peak power during propagation, and Fig.~\ref{fig:pulse_power_transition} showing the corresponding power profiles. Note that the behavior of these transitions is consistent with the nature of the unstable eigenvalues found for LAS and for the unstable MAS. Thus, the purely imaginary unstable eigenvalue of the LAS is in agreement with the abrupt transition, here observed from the LAS to MAS, and the two complex conjugated unstable eigenvalues of the MAS justify the oscillatory evolution of the peak of the MAS observed in Fig.~\ref{fig:peak_power_transition} around $z = 800~\text{m}$.

\begin{figure}[hbt!]
    \centering
    \subfigure[]{
    \centering\includegraphics[width = 0.48\textwidth, angle=0]{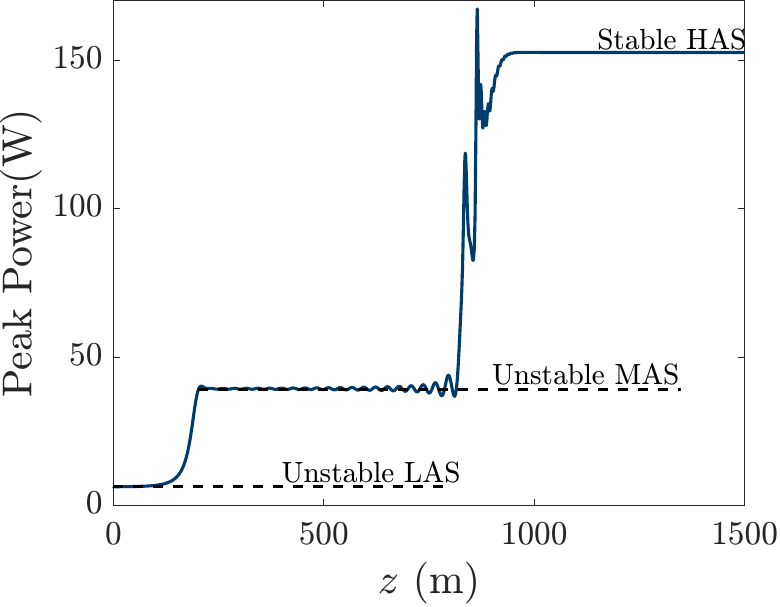}\label{fig:peak_power_transition}}
    \subfigure[]{
    \centering\includegraphics[width = 0.48\textwidth, angle = 0]{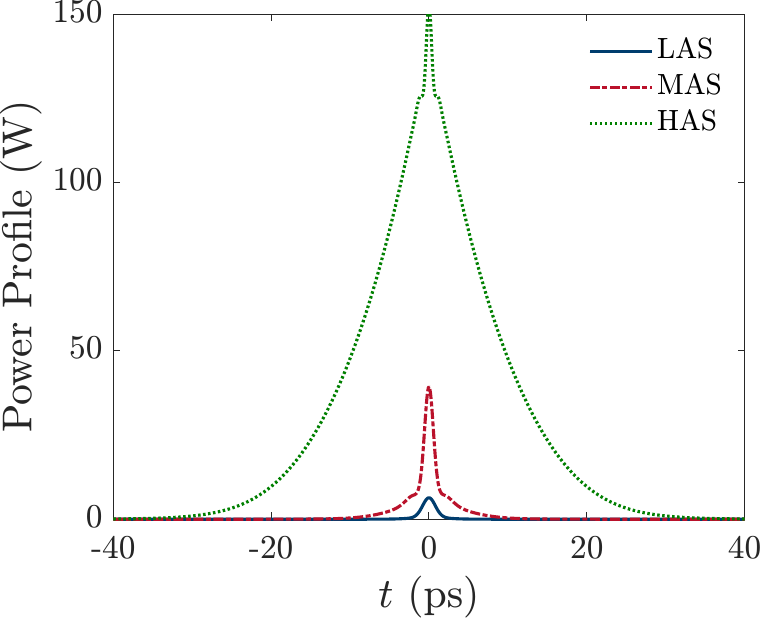}\label{fig:pulse_power_transition}}
    \caption{Evolution from an LAS solution to a HAS solution (passing through an MAS solution) for $g_0 = 1.47~\text{m}^{-1}$ and $T_2 = 100~\text{fs}$. \subref{fig:peak_power_transition} Evolution of the peak power in $z$ and \subref{fig:pulse_power_transition} power profiles of the three solutions.}
    \label{fig:solution_transition}
\end{figure}

\newpage
\subsection{Energy and Width} \label{sec:ew}

One point that has been raised about quartic solitons in mode-locked lasers is their ability to reach very high energy and very short temporal widths since their conservative versions have energy that scales inversely with the width cubed. In a previous work \cite{malheiro23}, we have shown that this scaling was only valid for this model for some of the solitons that exist for negative 4OD. To understand how energy and width depend on the parameters $(g_0,T_2)$, we used all the stable MAS and HAS that were obtained to produce Figs.~\ref{MAS_Ew} and \ref{HAS_Ew} showing energy ($E$) versus width ($\tau$). Each type of solution presents a different behavior. For HAS, the logarithms of energy and width fall in a single line giving the relation $E=16\,\tau^{2.0}$ (for energies in pJ and widths in ps) as presented in Fig.~\ref{HAS_Ew}. This energy-width relation is in agreement with the one estimated using the approximated profile given by Eq.~(\ref{eq:HAS_fit}). 
Regarding the energy-width relation for MAS, there are different curves for each $g_0$ and each curve does not follow a power law. Moreover, the energy may decrease or increase with the width depending on the region of the parameter space, such that, for lower $T_2$ the energy tends do decrease with width but, for higher $T_2$, the energy tends to increase with width.  These results for MAS do not follow the $E\propto \tau^{-3}$ that was observed for the CGLE solutions in \cite{wang24}.

\begin{figure}[hbt!]
    \subfigure[]{
        \centering\includegraphics[height=0.25\textheight]{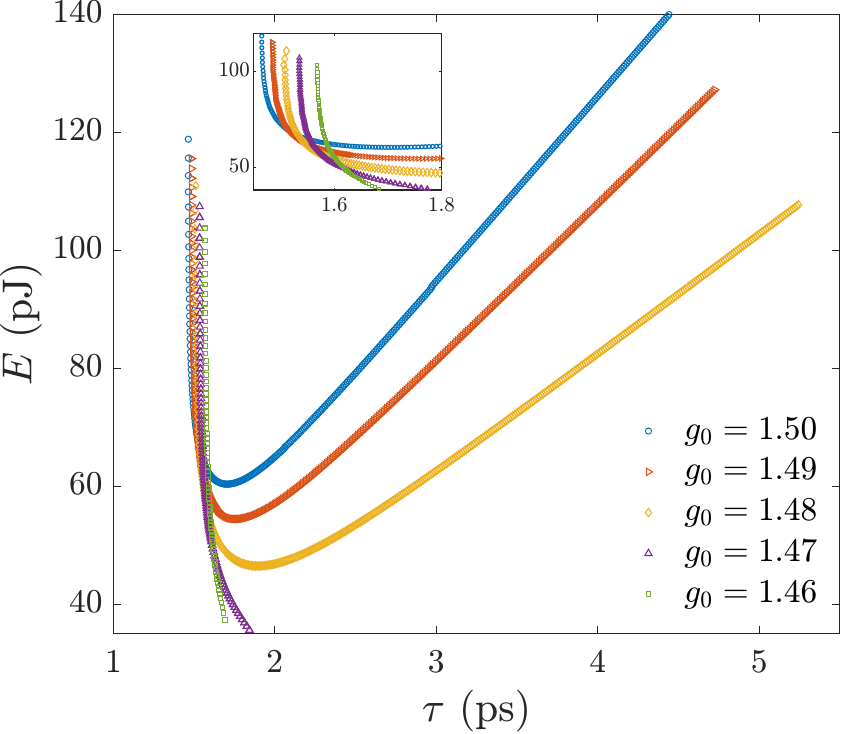}
        \label{MAS_Ew}}
    \subfigure[]{
        \centering\includegraphics[height=0.25\textheight]{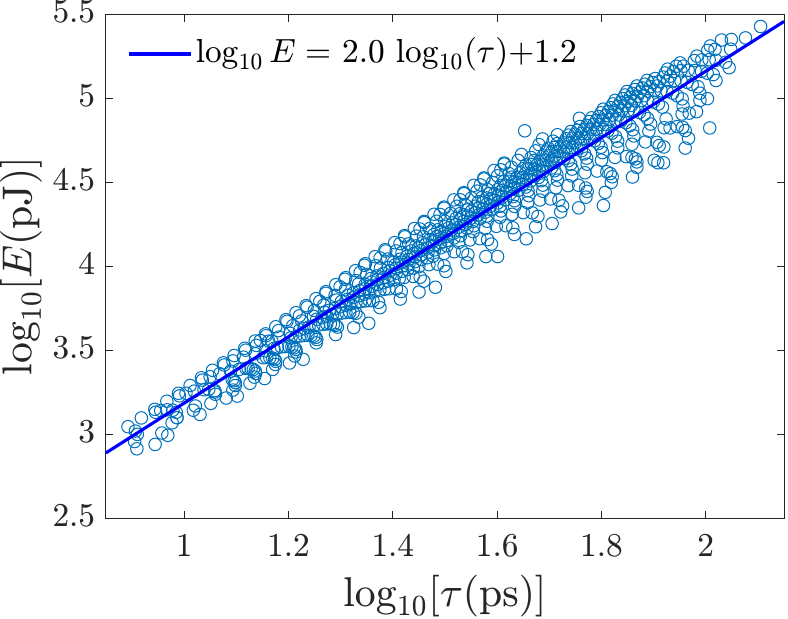}
        \label{HAS_Ew}}
\caption{\label{Ew} Energy versus temporal width for \subref{MAS_Ew} MAS and several $g_0$ (m$^{-1}$) values as indicated in the figure and for all the obtained HAS solutions \subref{HAS_Ew} in base 10 logarithms. The line in \subref{HAS_Ew} is a linear fit to the points whose equation is also on the graph.} 
\end{figure}

\section{Modulation instability} \label{sec:modulation_instability}
The existence of solitons has long been associated with modulational instability \cite{yuen75,tai86,Soto-Crespo02}. To assess the possibility of soliton generation from modulational instability in our model, we first found the continuous wave (cw) solution of Eq.~(\ref{PDE2}) and wrote down its stability equations. Thus, the cw solution of Eq.~(\ref{PDE2}) is given by
\begin{equation}
    q(T,Z)=A \exp(i\sigma Z),
    \label{eq:cw_sol}
\end{equation}
with $A=0$ and $\sigma=0$ (the zero homogeneous solution) or $A=\sqrt{1/\alpha\rho}$ and $\sigma=1/\alpha\rho$. Similarly to the analysis made for solitons, we may perturb the cw solution in the form
$q(T,Z)=\left[A+\eta(T,Z)\right] \exp(i\sigma Z)$, with $|\eta|\ll A$. Then, to first order in $\eta$, we have:
\begin{equation}
    i\eta_Z + \textbf{K}_{11}(A)\eta + \textbf{K}_{12}(A)\eta^\ast = 0.
    \label{eq:perturb_evolution2}
\end{equation}
Assuming that  $\eta(T,Z)=u(T)e^{i\lambda Z}+x^*(T)e^{-i\lambda^*Z}$, where $u(T)$ and $x(T)$ are periodic functions with frequency $\Omega$, that is, $u(T)=u_0e^{i\Omega T}$ and $x(T)=x_0e^{i\Omega T}$ and by separating the terms in $e^{i\lambda Z}$ and $e^{-i\lambda^*Z}$, we find that $u_0$ and $x_0$ satisfy the eigenvalue equation 
\begin{equation}
\begin{bmatrix}
A_{11} & A_{12}  \\
-A^*_{12} & -A^*_{11}
\end{bmatrix}\;\begin{bmatrix} u_0 \\ x_0\end{bmatrix}=\lambda\; \begin{bmatrix} u_0 \\ x_0\end{bmatrix}
\label{MI_eq}
\end{equation}
with 
\begin{equation*}
A_{11}=\frac{D_4}{24}\Omega^4 + i\Omega^2 - \sigma - i\alpha + 2A^2 + i\frac{1+\alpha}{\left(1+\rho A^2\right)^2},
\label{A11}
\end{equation*}

\begin{equation*}
A_{12}=A^2 - i\frac{\left(1+\alpha\right)\rho A^2}{\left(1+\rho A^2\right)^2}.
\label{A12}
\end{equation*}

The eigenvalues $\lambda$ are given by

\begin{equation}
\lambda= i\Im{A_{11}}  \pm \sqrt{\left(\Re{A_{11}}\right)^2-|A_{12}|^2}.
\end{equation}
For the zero homogeneous solution with $A=\sigma=0$, the eigenvalues are given by $\lambda=\pm D_4\Omega^4/24+i\Omega^2+i\alpha$ which implies stability if $\alpha>0$. For the nonzero solution with $A=\sqrt{1/\alpha\rho}$ and $\sigma=1/\alpha\rho$, the eigenvalues are given by
\begin{equation}
\lambda=-i\frac{\alpha}{1+\alpha}+i\Omega^2\pm\sqrt{\frac{D_4^2}{576}\Omega^8+\frac{D_4}{12\alpha\rho}\Omega^4-\frac{\alpha^2}{(1+\alpha)^2}}   
\label{lambda}
\end{equation}
This equation indicates that, for lower $|\Omega|$ the two eigenvalues will be purely imaginary, but for $|\Omega|>\Omega_c$, with $\Omega_c$ given by
$$\Omega_c^4=\frac{24}{\alpha\rho D_4}\left(-1+\sqrt{1+\alpha^4\rho^2/(1+\alpha)^2}\right),$$
the two eigenvalues have symmetrical real parts and an imaginary part equal to $-\alpha/ (1+\alpha)+\Omega^2$. Moreover, the lower imaginary part is achieved for $\Omega=0$ and is equal to $-2\alpha/(1+\alpha)$. Since the cw solution is modulational unstable whenever $\lambda_i<0$, we find that there is always a $\lambda_i<0$ as long as $\alpha>0$, i.e., for $g_0>k_\text{OC}/L$, which was always the case whenever we found solitons. Moreover, the maximum growth is
$$g_\text{max}=-\lambda_i^\text{min}=\frac{2\alpha}{1+\alpha},\quad\text{at}\quad \Omega=0.$$
For $\Omega\rightarrow\infty$, the imaginary part of the eigenvalues behave as $\lambda_i\rightarrow \Omega^2$, which corresponds to stable modes. 
Fig.~\ref{MI_lambdas} shows a typical graph for the imaginary parts of $\lambda$ as function of $\Omega$ for $\alpha>0$.

\begin{figure}[hbt!]
        \centering
        \includegraphics[width=0.48\textwidth]{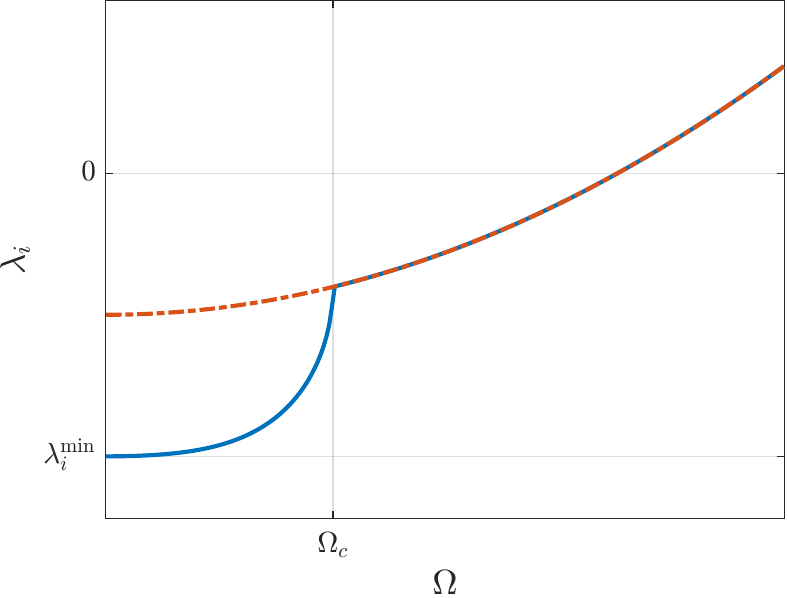}      
\caption{Typical imaginary parts of the modulational instability eigenvalues given by Eq.~(\ref{lambda}), for the nonzero homogeneous solution as function of modulational frequency $\Omega$ of the perturbation modes, considering $\alpha>0$.}\label{MI_lambdas}
\end{figure}

We have propagated the homogeneous solution slightly perturbed with noise, i.e., using an input like $q(T,0)=\sqrt{1/\alpha\rho}+0.01\mu
$, where  $\mu$ is a random variable with uniform distribution between 0 and 1, in an effort to understand if the modulational instability gives rise to solitons. 
For parameter values inside of the region of existence of stable solitons, we observe two different behaviors depending on the type of solitons that exist on that region. Thus, for parameters for which HAS exist, even if in coexistence with MAS, the perturbed homogeneous solution does not evolve to solitons and the typical evolution is shown in Fig.~\ref{MI-HAS}.
On the other hand, for parameters in the region where only MAS exist, the homogeneous solution evolves into a train of MAS. This type of evolution is shown in Fig.~\ref{MI-MAS}, where we may observe three pulses with the final form of the corresponding MAS for these parameters and another one still evolving to the final form.
Away from the parameter region of stable solitons, there are also two types of evolution. For $T_2$ below the lower boundary of HAS, the homogeneous solution evolves to a high irregular amplitude cw solution similar to the one in Fig.~\ref{MI-HAS} but, for $T_2$ above the higher boundary of MAS, the homogeneous solution decays to zero.

\begin{figure}[hbt!]
    \subfigure[]{
        \centering\includegraphics[width =0.48\textwidth, angle=0]{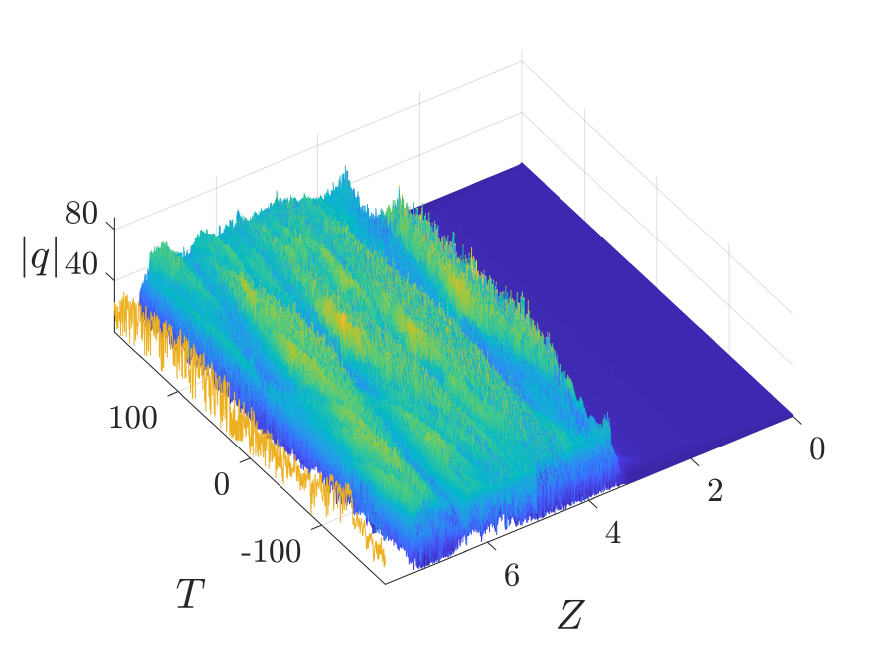}
        \label{MI-HAS}}
    \subfigure[]{
        \centering\includegraphics[width =0.48\textwidth, angle = 0]{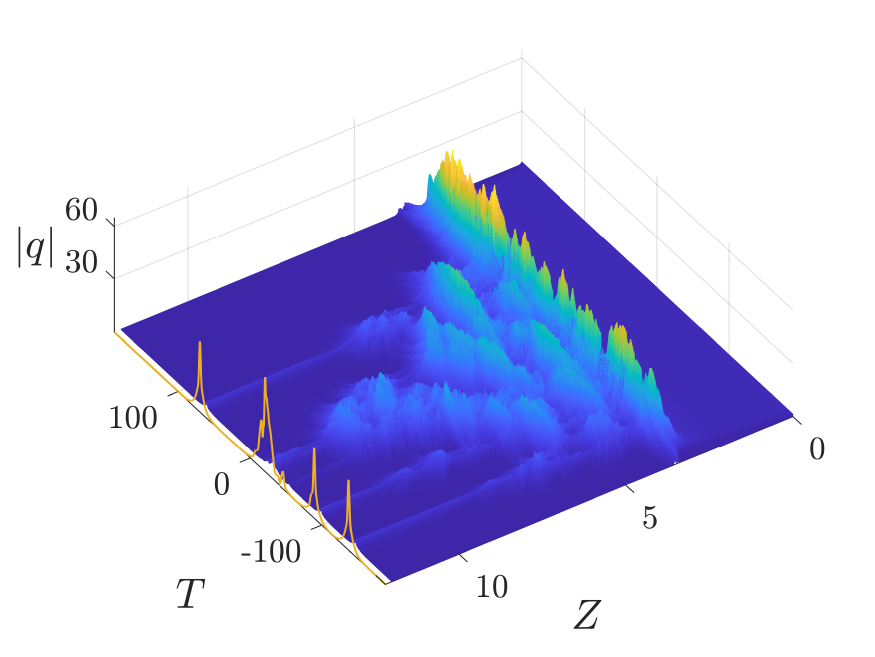}
        \label{MI-MAS}}
\caption{\label{MI} Evolution of a perturbed homogeneous solution for \subref{MI-HAS} $g_0=1.46$~m$^{-1}$ and $T_2=80$~fs, \subref{MI-MAS} for $g_0=1.46$~m$^{-1}$ and $T_2=120$~fs.The final line in each graph corresponds to the final $|q|$ output, being at the same scale as the 3D plot in the case of \subref{MI-HAS} but multiplied by 10 in the case of \subref{MI-MAS} in order to become visible.}
\end{figure}

\section{\label{2&3_influence}Effect of second and third order dispersion}
In practical realizations, it may be difficult to create conditions for effective zero second and third order dispersion. Here, we briefly analyze the effect of $\beta_2$ and $\beta_3$ to MAS and HAS.

Considering $T_2$ and $g_0$ inside the region of existence and stability of HAS, namely $T_2=60$~fs and $g_0=1.4$~m$^{-1}$, we observed the evolution of the corresponding HAS in case $\beta_2$ is nonzero. 
The evolution tends to new stationary profiles that are shown in Fig.~\ref{HAS_beta2} for $\beta_2=\pm 0.1$~ps$^2$m$^{-1}$. These results show that positive $\beta_2$ increases the amplitude and energy of the pulse, eliminating the thin spike of the HAS. On the other hand, negative $\beta_2$ decreases the amplitude and energy and increases the structure of the top spike. A similar approach was used to understand the effect of $\beta_3$. The final stationary soliton profiles are shown in Fig.~\ref{HAS_beta3} side by side with the pure quartic HAS for the same parameters as in the above paragraph but with $\beta_2=0$ and $\beta_3=\pm 0.05$~ps$^3$m$^{-1}$, which is close to the maximum allowed $\beta_3$ that preserves stationary evolution of the pulse. The profiles are asymmetrical and have slightly higher amplitude than the pure quartic HAS. In the figure, the peak position of the nonzero $\beta_3$ profiles are to the left and right of the HAS peak, depending on the sign of $\beta_3$. In the presence of $\beta_3$, the pulse deforms asymmetrically and the effective group velocity differs from the assumed group velocity calculated for the central wavelength. Thus, in the time referential used in equation (\ref{PDE}) that travels with the latter velocity, the pulses are displaced from the initial temporal position by $t(z)=t(0)+vz$, with $v$ being the shift in velocity introduced by $\beta_3$. Note that the asymmetry and velocity are identical but have opposite directions depending on the sign of $\beta_3$.

\begin{figure}[hbt!]
        \centering
        \subfigure{\label{HAS_beta2}\includegraphics[width=0.48\textwidth]{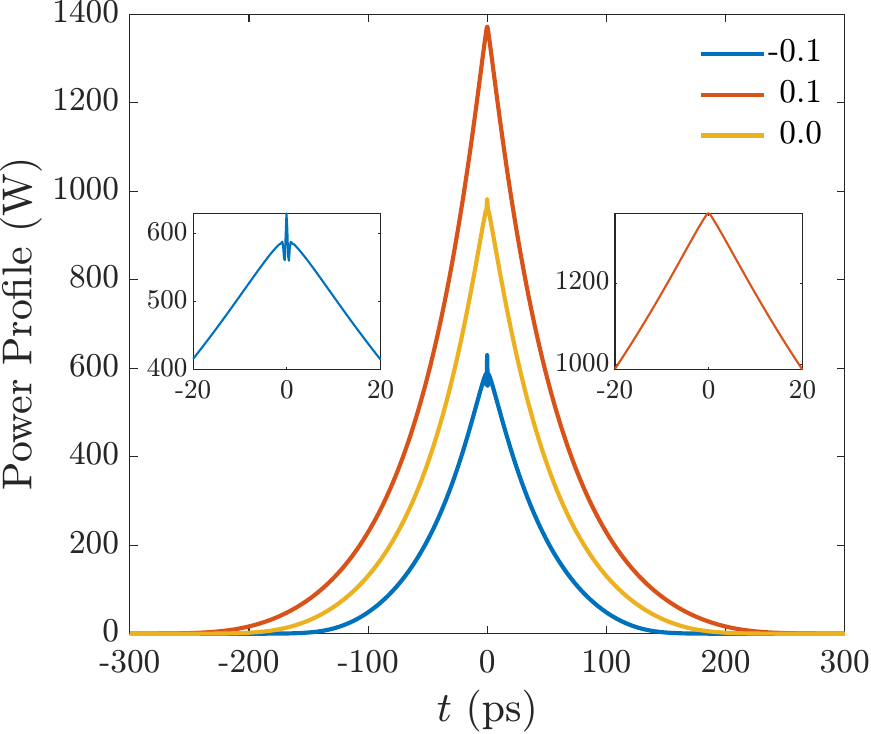} } 
        \subfigure{\label{HAS_beta3}\includegraphics[width=0.48\textwidth]{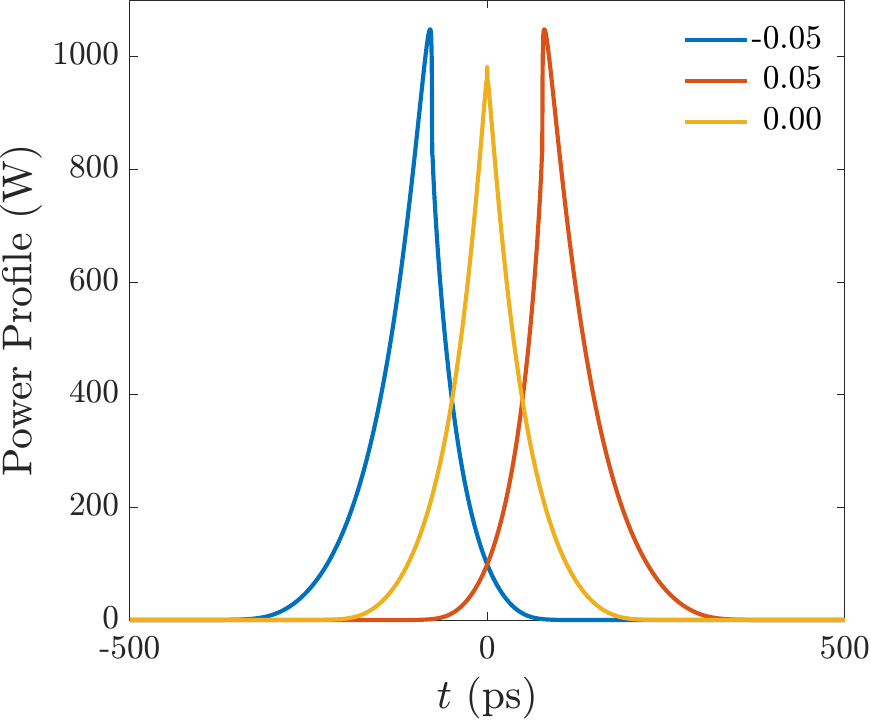} }  
\caption{Power profiles of the stationary solitons in the region of stable HAS ($T_2=60$~fs and $g_0=1.4$~m$^{-1}$) for \subref{HAS_beta2} $\beta_2=\pm 0.1$ ~ps$^2$m$^{-1}$ and \subref{HAS_beta3} $\beta_3=\pm 0.05$ ~ps$^3$m$^{-1}$ as indicated in the legends. The profile for $\beta_2=0$ is added for comparison. A zoom of the peaks is shown, in \subref{HAS_beta2} for both cases of nonzero $\beta_2$.}\label{HAS_beta23}
\end{figure}

To further assess the resilience of quartic solitons in the presence of other dispersion orders, we obtained the stationary solitons in the HAS regions in the presence of second, third and fourth dispersion orders and the power profiles are shown in Fig.~\ref{HAS_beta2_beta3}. We note that the asymmetry is present due to the third order dispersion and the peak amplitudes are in agreement with previous results (Fig~\ref{HAS_beta2}) for negative and positive $\beta_2$.

\begin{figure}[hbt!]
        \centering
        \includegraphics[width=0.5\textwidth]{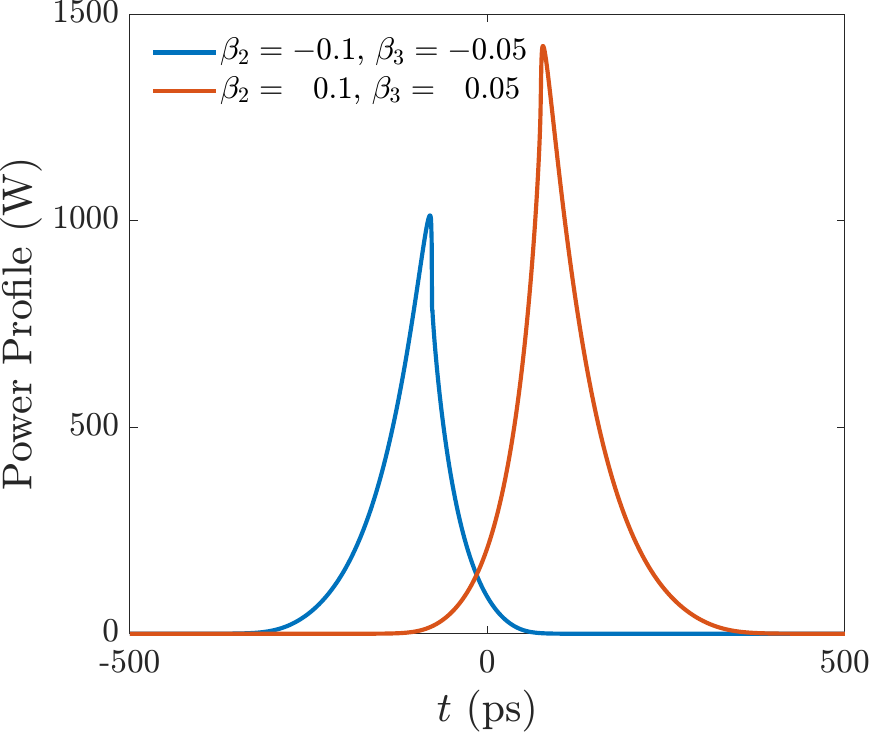}      
\caption{Power profiles of the stationary solitons in the region of stable HAS ($T_2=60$~fs and $g_0=1.4$~m$^{-1}$) in the presence of both $\beta_2$ and $\beta_3$ as indicated in the legend in ps$^2$m$^{-1}$ and ps$^3$m$^{-1}$, respectively.}\label{HAS_beta2_beta3}
\end{figure}

Turning the attention to the effect of $\beta_2$ and $\beta_3$ on the region of existence and stability of MAS, we have considered $T_2=150$~fs and $g_0=1.48$~m$^{-1}$ and observed the evolution of the corresponding MAS. In Fig.~\ref{MAS_beta2}, three stationary profiles are shown, the MAS itself and the ones for $\beta_2=\pm 0.1$ ps$^2$m$^{-1}$. A negative $\beta_2$ increases the amplitude and induces two side humps. On the other hand, a positive $\beta_2$ clearly increases amplitude, width and energy and produces a regular bell-shaped pulse. The effect of $\beta_3$ on the region of stable MAS is similar to the effect already described for solutions on the region of HAS, i.e., the amplitude of the pulses are slightly higher than the corresponding MAS but asymmetric and showing a one-side pedestal. They also have a nonzero velocity. However, a difference relative to the effect on the HAS region is on the maximum $|\beta_3|$ that supports stable propagation which, in this case, is higher. We obtained stable propagation for $|\beta_3|$ as high as 0.7~ps$^3$m$^{-1}$, for which the peak power is 10 times higher than the corresponding MAS for approximately the same width. 

\begin{figure}[hbt!]
        \centering
        \subfigure{\label{MAS_beta2}\includegraphics[width=0.48\textwidth]{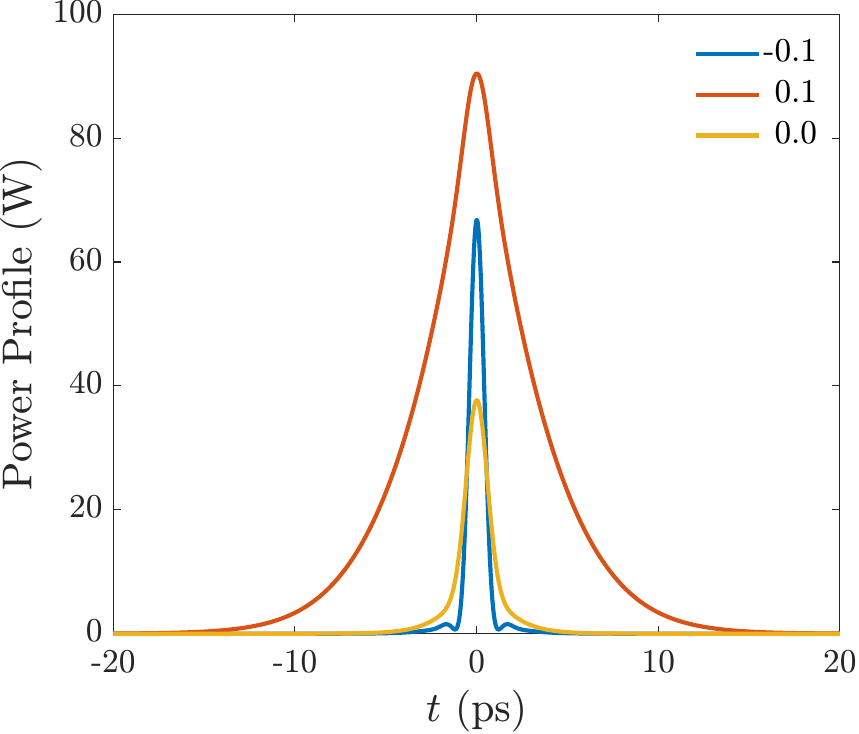} } 
         \subfigure{\label{MAS_beta3}\includegraphics[width=0.48\textwidth]{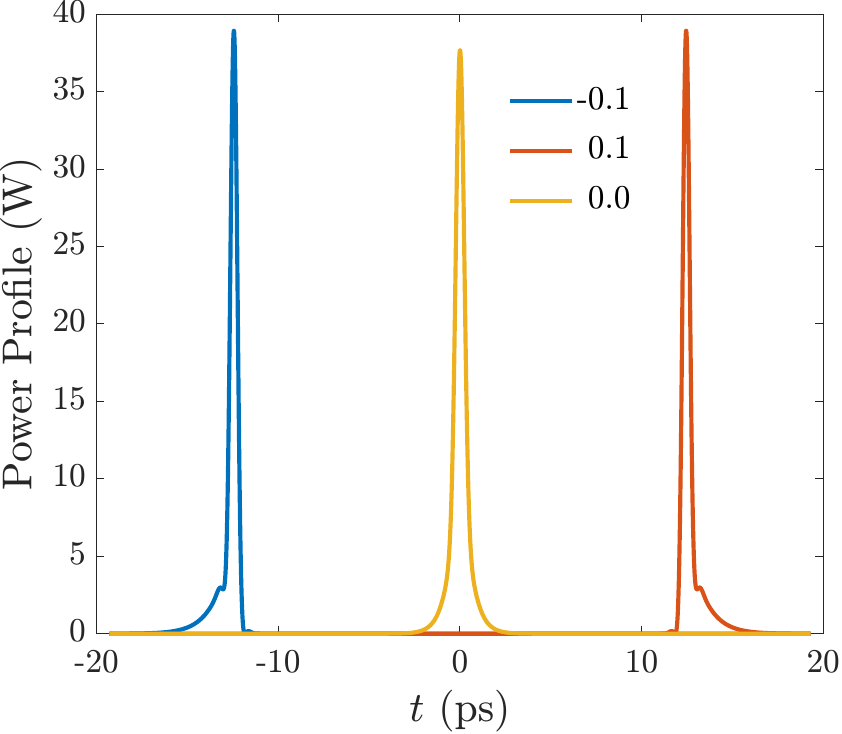}  }
\caption{Power profiles of the stationary solitons in the region of stable MAS ($T_2=150$~fs and $g_0=1.48$~m$^{-1}$) for \subref{MAS_beta2} $\beta_2=\pm 0.1$ ~ps$^2$m$^{-1}$ and \subref{MAS_beta3} for $\beta_3=\pm 0.1$ ~ps$^3$m$^{-1}$ as indicated in the legends. The profile for $\beta_2=0$ is added for comparison. }\label{MAS_beta23}
\end{figure}

Finally, in Fig.~\ref{MAS_beta2_beta3} we present the stationary profiles for the solutions in the region of MAS with both nonzero $\beta_2$ and $\beta_3$. The profiles exhibit the characteristics already shown for $\beta_2$ and $\beta_3$ separately, namely, the asymmetry characteristic of $\beta_3$ and the profile shapes associated with the negative and positive $\beta_2$. 

\begin{figure}[hbt!]
        \centering
        \includegraphics[width=0.5\textwidth]{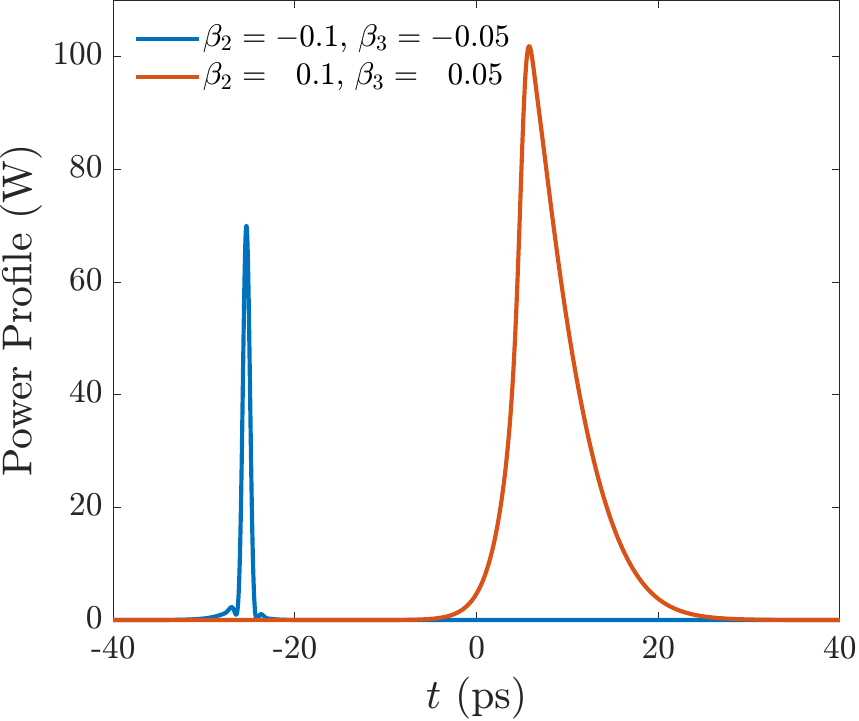}      
\caption{Power profiles of the stationary solitons in the region of stable MAS ($T_2=150$~fs and $g_0=1.48$~m$^{-1}$) in the presence of both $\beta_2$ and $\beta_3$ as indicated in the legend in ps$^2$m$^{-1}$ and ps$^3$m$^{-1}$, respectively.}\label{MAS_beta2_beta3}
\end{figure}

We finish this section with two final remarks. The goal of those tests were to show that the MAS and HAS obtained for a purely positive 4OD medium are resistant to the presence of both 2OD and 3OD. Nevertheless, we have realized that if the magnitude of $\beta_2$ and $\beta_3$ are not neglectful the characteristics of the solutions are, in some cases, reasonably different from the MAS and the HAS, possible with practical interesting characteristics and laws that are worth of a more detailed study. Moreover, those results are not unexpected since we know that the balance that permits dissipative solitons does not only rely on dispersion and nonlinear effect but also on loss and gain.

\section{Conclusions}\label{sec:conclusion}
An in-depth study about soliton solutions of a distributed model for a mode-locked laser, dominated by positive fourth order dispersion, was carried out. Some of the types of solutions have no practical importance since they are always unstable. Two types of solutions exist and are stable for a relatively large regions of linear gain and spectral filtering parameters. They even coexist and are stable in a particular region of parameters, thus, the model presents effects of bistability, which could have application in optical logic gates. These two kinds of solutions differ in their energy-width relation. The energy of the high amplitude solution scales with the square of the FWHM measured from the thin spike on the top of the pulse that is characteristic of this type of solution. On the other hand, the energy of the medium amplitude solution has no fixed relation with the temporal width, it may increase or decrease with the temporal width. For lower spectral filtering and higher linear gain parameters, this type of solution has the advantage of being energetic and short. Moreover, this same type of solution, i.e, the MAS may be self excited through modulational instability which corresponds to an advantage for the development of self-starting pulsed lasers. Both MAS and HAS do survive in the presence of second and third order dispersion contributions. The preliminary results also reveal that these lower order dispersion contributions together with 4OD may give rise to new types of solutions that will be worth studying in future works. All these results should be interesting to both theoretical and experimental communities studying and implementing mode-locked lasers. In particular, they will be useful for identifying laser implementations that produce pulses with the desired characteristics—either by selecting setups with similar parameters to those used here, or by using the adimensionalized variables and parameters, also introduced here, to predict pulse behavior for other achievable configurations.


\begin{acknowledgments}
This work has received funding from the Fundação para a Ciência e a Tecnologia (FCT) within the Projects LA/P/0037/2020,
UIDB/50014/2020, UIDB/50025/2020, UIDP/50025/2020, and PTDC/FIS-OUT/3882/2020. D. Malheiro acknowledges a PhD fellowship from FCT with code 2024.01557.BD.
\end{acknowledgments}

%

\end{document}